\documentclass[aps,pra,twocolumn,showpacs,notitlepage,superscriptaddress,letterpaper]{revtex4-2}
\usepackage{graphicx, amsmath, amssymb, amsfonts, pifont, bm, cancel}
\usepackage[usenames]{color}
\usepackage{subfigure}
\usepackage[normalem]{ulem} 
\usepackage{circuitikz}
\usepackage{hyperref}
\usepackage{upgreek}
\usepackage{outlines}
\usepackage{enumitem}
\usepackage{blkarray}

\setenumerate[1]{label=\Roman*.}
\setenumerate[2]{label=\Alph*.}
\setenumerate[3]{label=\roman*.}
\setenumerate[4]{label=\alph*.}

\definecolor{MyDarkBlue}{rgb}{0,0,1}
\definecolor{darkgreen}{rgb}{0.0, 0.5, 0.0}

\newcommand{\beq}{\begin{equation}}
\newcommand{\eeq}{\end{equation}}
\newcommand{\bel}{\begin{align*}}
\newcommand{\tamam}{\end{align*}}

\newcommand{\ket}[1]{|#1\rangle}

\newcommand{\beqa}{\begin{eqnarray}}             
\newcommand{\eeqa}{\end{eqnarray}}               
\newcommand{\bra}[1]{\langle#1\vert}                 

\newcommand{\GammaoneD}{\Gamma_{\text{1D}}}

\usepackage{physics}

\usepackage{physics}

\begin{document}

\title{Fast Unconditional Reset and Leakage Reduction of a Tunable Superconducting Qubit via an Engineered Dissipative Bath
}

\author{Gihwan Kim}
\thanks{These authors contributed equally}
\affiliation{Kavli Nanoscience Institute and Thomas J. Watson, Sr., Laboratory of Applied Physics, California Institute of Technology, Pasadena, California 91125, USA.}
\affiliation{Institute for Quantum Information and Matter, California Institute of Technology, Pasadena, California 91125, USA.}

\author{Andreas Butler}
\thanks{These authors contributed equally}
\affiliation{Kavli Nanoscience Institute and Thomas J. Watson, Sr., Laboratory of Applied Physics, California Institute of Technology, Pasadena, California 91125, USA.}
\affiliation{Institute for Quantum Information and Matter, California Institute of Technology, Pasadena, California 91125, USA.}

\author{Vinicius S. Ferreira}
\thanks{Current Address: Google Quantum AI, Santa Barbara, CA}
\affiliation{Kavli Nanoscience Institute and Thomas J. Watson, Sr., Laboratory of Applied Physics, California Institute of Technology, Pasadena, California 91125, USA.}
\affiliation{Institute for Quantum Information and Matter, California Institute of Technology, Pasadena, California 91125, USA.}
\
\author{Xueyue (Sherry) Zhang}
\thanks{Current Address: Miller Institute, University of California Berkeley, Berkeley, CA 94720.} 

\affiliation{Kavli Nanoscience Institute and Thomas J. Watson, Sr., Laboratory of Applied Physics, California Institute of Technology, Pasadena, California 91125, USA.}
\affiliation{Institute for Quantum Information and Matter, California Institute of Technology, Pasadena, California 91125, USA.}
\author{Alex Hadley}
\affiliation{Kavli Nanoscience Institute and Thomas J. Watson, Sr., Laboratory of Applied Physics, California Institute of Technology, Pasadena, California 91125, USA.}
\author{Eunjong Kim}
\affiliation{Department of Physics \& Astronomy, Seoul National University, Seoul 08826, Korea} 
\affiliation{Institute of Applied Physics, Seoul National University, Seoul 08826, Korea}

\author{Oskar~Painter}
\email{opainter@caltech.edu}
\homepage{http://painterlab.caltech.edu}
\affiliation{Kavli Nanoscience Institute and Thomas J. Watson, Sr., Laboratory of Applied Physics, California Institute of Technology, Pasadena, California 91125, USA.}
\affiliation{Institute for Quantum Information and Matter, California Institute of Technology, Pasadena, California 91125, USA.}

\date{\today}
\begin{abstract}
Rapid and accurate initialization of qubits, reset, is a crucial building block for various tasks in quantum information processing, such as quantum error-correction and estimation of statistics of noisy quantum devices with many qubits. We demonstrate unconditional reset of a frequency-tunable transmon qubit that simultaneously resets multiple excited states by utilizing a metamaterial waveguide engineered to provide a cold bath over a wide spectral range, while providing strong protection against Purcell decay of the qubit. We report reset error below 0.13\% (0.16\%) when prepared in the first (second) excited state of the transmon within 88ns. Additionally, through the sharp roll-off in the density of states of the metamaterial waveguide, we implement a leakage reduction unit that selectively resets the transmon's second excited state to 0.285(3)\% residual population within 44ns while acting trivially in the computational subspace as an identity operation that preserves encoded information with an infidelity of 0.72(1)\%.
\end{abstract}
\maketitle

\clearpage

\label{Section:Introduction}
\textit{Introduction.}---Fast and high-fidelity reset of qubits into fiducial states is a necessary capability of quantum processors designed for demanding quantum information and computation tasks~\cite{divincenzo2000physical}. In the context of quantum error-correction, high-fidelity qubit reset minimizes faulty state preparation of auxillary qubits, and rapid qubit reset reduces the idling error experienced by spectator qubits~\cite{ng2011dynamical, reed2012realization, chen2021exponential, krinner2022realizing, google2023suppressing, geher2024reset}. In addition, operations capable of reseting leakage states into the computational subspace provide the benefit of converting uncorrectable leakage errors into correctable errors~\cite{mcewen2021removing, miao2023overcoming, battistel2021hardware, marques2023allmicrowaveleakage, lacroix2023fast}. More generally, fast and accurate qubit reset improves experimental repetition rates and state preparation fidelities, bringing the measurement times of higher-order statistics of noisy many-qubit devices within a feasible range. 

In superconducting qubit systems, reset schemes that use projective single-shot measurement followed by conditional single-qubit control can realize reset within a few hundred nanoseconds~\cite{riste2012initialization, riste2012feedback, johnson2012heralded, campagne2013persistent}. However, measurement-based conditional reset is limited in speed by inevitable temporal overhead in post-processing, and limited in accuracy by readout infidelity, for example due to measurement-induced state transitions~\cite{sank2016measurement, khezri2023measurement}. 

More recently, state-of-the-art unconditional transmon qubit reset has been achieved by activating resonant exchange coupling between a transmon and a damped auxiliary mode such as a readout resonator, and allowing the system to decay to its collective ground state~\cite{reed2010fast, geerlings2013demonstrating, magnard2018fast, mcewen2021removing, zhou2021rapid, sunada2022fast, chen2024fast}. Such reset protocols prepare the ground state independent of the prior state of the qubit, and thus are not limited by readout time and fidelity. In addition, unconditional reset schemes help reduce the impact of state misclassification errors on fault-tolerant logical operations~\cite{geher2024reset}. Still, the speed of these reset schemes is limited by the decay rate of the auxiliary mode, which is itself constrained by the need to suppress Purcell decay and, in the case where the mode belongs to the qubit's readout resonator, optimize readout fidelity~\cite{purcell1946spontaneous, jeffrey2014fast, zhou2021rapid, mcewen2021removing}. Furthermore, additional driving or complicated pulse shaping is required for each additional excited state to be reset due to the narrow-banded nature of the auxiliary modes. 

In this work, we go beyond the state of the art by implementing fast, unconditional reset of the first two excited states of a frequency-tunable transmon qubit by coupling it to a broadband metamaterial waveguide (MMWG) strongly damped to a cold environment~\cite{mirhosseini2018superconducting, kim2021quantum, ferreira2021collapse, zhang2023superconducting, ferreira2024deterministic}. The MMWG provides stronger protection against Purcell decay compared to single-pole Purcell filters, thanks to the sharp extinction of its density of states (DOS) outside of its passband. This enables significantly faster reset than in previous works by alleviating the tradeoff between large achievable damping and idling qubit Purcell relaxation. The bandwidth of the MMWG can be made large relative to the transmon anharmonicity, allowing for simultaneous reset of multiple excited states. Additionally, bandedges with rolloff that is steep relative to the transmon anharmonicity provide a means for implementing so-called leakage reduction units (LRUs)~\cite{battistel2021hardware}, which selectively reset the second excited state of the transmon without corrupting the information encoded in the computational subspace. Leveraging these capabilities of the MMWG combined with dynamical activation of the interaction via parametric flux modulation~\cite{beaudoin2012firstorder, strand2013firstorder, silveri2017quantum}, we demonstrate simultaneous reset of a transmon's three lowest energy eigenstates and an LRU that selectively resets a transmon's second excited state. 

\begin{figure}[tbp]
\centering
\includegraphics[width = \columnwidth]{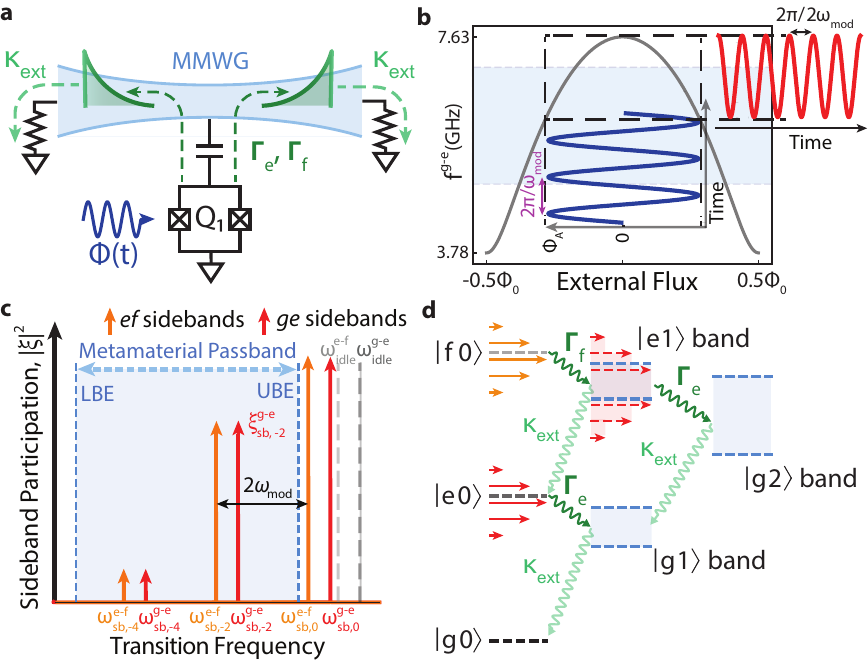}
\caption{\textbf{Schematic of the reset protocol.} \textbf{a} Illustration of reset of Q\textsubscript{1}. External flux pulse $\Phi(t)$ applied to the SQUID loop of Q\textsubscript{1} causes bidirectional emission of excitation from Q\textsubscript{1} into the MMWG, which is damped to its external environment. \textbf{b} Tuning curve of the flux-tunable transmon qubit. $\Phi_0$ is the magnetic flux quantum. Parking the qubit at the upper sweet-spot, where the qubit is first-order insensitive to the external flux, a flux modulation tone of frequency $\omega_{\textrm{mod}}$ (dark blue curve) modulates the qubit frequency at a frequency $2\omega_{\textrm{mod}}$ (red curve). The region shaded in pale blue denotes the extent of the MMWG passband. \textbf{c} Sideband picture of the Q\textsubscript{1} during reset operation. Gray dashed lines represent idling g-e and e-f transition frequencies of Q\textsubscript{1}, while colored lines indicate sideband frequencies with parametric modulation at $\omega_{\textrm{mod}}$ turned on. \textbf{d} Energy level diagram showing flow of reset for different states. The notation $\ket{s\, n}$ is shorthand for Q\textsubscript{1} in state $\ket{s}$ and $n$ photons being in the waveguide. Red and gold arrows denote modulation sidebands. Note states of nonzero MMWG photon number live in a band of closely spaced energy levels, represented by the the semi-opaque blue bands. The transition frequencies of the states in these bands can also be modulated when they include excitations in the qubit, as illustrated by the semi-opaque red rectangles representing the set of sidebands of the $\ket{e\, 1}$ band, because they depend on qubit frequency.} 
\label{fig:ResetScheme}
\end{figure}

\label{Section:Results}
\textit{Results.}---We perform our measurements on a frequency-tunable transmon qubit (Q\textsubscript{1}) on the same device used in Ref.~\cite{zhang2023superconducting}, biased at its upper sweet-spot (USS) with a g-e transition frequency $\omega_{ge} / 2\pi = 7.63$ GHz and an anharmonicity $\eta/2\pi \equiv (\omega_{ef} - \omega_{ge})/2\pi = -179$ MHz. The qubit has a dedicated microwave drive line for XY control and a flux line for slow and fast frequency control. 

Q\textsubscript{1} is capacitively coupled to a single unit cell of a MMWG comprising a 1D coupled array of 52 lumped-element microwave resonators, as schematically shown in Fig.~\ref{fig:ResetScheme}a. Such a structure can be modeled as a set of harmonic modes with a finite passband with sharp bandedges, and allows rapid photon emission from coupled modes~\cite{calajo2016atom, gonzalez2017markovian, ferreira2021collapse}. At either end of the MMWG is a tapering section consisting of four resonators that impedance-match the MMWG modes to 50 $\Omega$ transmission lines serving as input and outputs to the waveguide, providing strong damping to its environment across a wide and flat passband from approximately 5 GHz to 7 GHz. We measure Q\textsubscript{1} using standard dispersive readout techniques, with the MMWG serving as a Purcell filter waveguide for readout. Because of the need to estimate small residual excitation population for characterizing qubit reset on the order of the thermal population of the MMWG, achieving sufficiently high sensitivity in single-shot readout is crucial. More details of device parameters, readout, and thermal population of the MMWG can be found in Appendix \ref{App:Device} - \ref{App:ThermalPop}. 

Our qubit reset protocol is based on dynamically activating an exchange interaction between Q\textsubscript{1} and the MMWG modes. When Q\textsubscript{1} is idling, detuned above the MMWG passband, the rapid extinction of the MMWG DOS outside its bandedges strongly protects Q\textsubscript{1} from Purcell decay through the 50$\Omega$ ports of the MMWG. To turn on an exchange interaction, we apply a single-tone flux modulation pulse to Q\textsubscript{1}'s SQUID loop that activates a resonance condition between Q\textsubscript{1} and the MMWG passband and rapidly converts this Purcell decay protection into Purcell decay enhancement, inducing qubit reset. 

As illustrated in Fig. \ref{fig:ResetScheme}b, c, when Q\textsubscript{1} is biased at its USS, a flux modulation pulse of frequency $\omega_{\textrm{mod}}$ and amplitude $\Phi_m$ grows sidebands separated by even integer multiples ($\pm n 2\omega_{\textrm{mod}}$) from the time-averaged transition frequencies. Note that the negative convexity of the qubit's tuning curve at its bias point, shifts the time-averaged transition frequencies of Q\textsubscript{1} downwards. 

Sidebands of the transmon that fall within the passband of the MMWG are coupled to an effective continuum of modes and mediate radiative emission into the MMWG, as illustrated in the energy level diagram of Fig. \ref{fig:ResetScheme}c and d. The rate of emission of the transmon is roughly set by Fermi's golden rule involving the DOS of the MMWG and the transmon-MMWG coupling matrix element, which is determined by the bare coupling between Q\textsubscript{1} and its unit cell, and the participation of the sidebands. Multiple excited states of the transmon can be simultaneously reset by the same mechanism, thanks to the large bandwidth of the MMWG as shown in Fig. \ref{fig:ResetScheme}c and d, where the emission processes for states up to the second-excited state are illustrated. 

Emitted photons leave the waveguide at an effective damping rate $\kappa_{\textrm{ext}}$ determined by the finite reflectivity of the MMWG tapers and a round trip time of $\approx$ 14ns, estimated from the group velocity and the number of unit cells in the MMWG~\cite{ferreira2021collapse}. Note that full reset of both qubit and waveguide is important to avoid subsequent qubit control errors caused by photons remaining in the waveguide potentially inducing population revival or dephasing~\cite{ferreira2021collapse, ferreira2024deterministic}.

We have chosen parameteric modulation in order to avoid shelving and retrieval of excitation in and out of long-lived resonances whose frequencies collide with Q\textsubscript{1}'s frequency during flux tuning~\cite{shevchenko2010landau}. We find multiple such resonances, including two-level system defects (TLSs), and MMWG modes near the bandedges that are not sufficiently damped to output lines, one of which we measure to have a relaxation time of $\approx$ 1.7us. See Appendix \ref{App:DirectTuning} for details of reset with direct flux tuning and Landau-Zener-St\"uckelberg interference with long-lived resonances.

To calibrate a reset operation for at least the first two excited states of Q\textsubscript{1}, we first define a flux pulse $\Phi(t; \Phi_\text{A}, \omega_{\textrm{mod}}, \tau_\text{pulse}) = \Phi_\text{A } \mathcal{E}(t; \tau_\text{pulse}) \sin(\omega_{\textrm{mod}} t)$ parameterized with $\Phi_\text{A}$ (modulation amplitude), $\omega_\text{mod}$ (modulation frequency), and $\tau_\text{pulse}$ (total flux pulse duration). The envelope function $\mathcal{E}(t; \tau_\text{pulse})$ is a square pulse of duration $\tau_\text{pulse}$ with zero-amplitude buffers $\tau_\text{B} = 2$ ns at the start and end filtered with a Gaussian kernel of width $\sigma = 1$ ns~\cite{lacroix2023fast}. We perform a 3D sweep of the parameters $\Phi_\text{A}$, $\omega_{\textrm{mod}}$, and $\tau_\text{pulse}$ in order to characterize the reset at each point. 

\begin{figure}[tbp]
\centering
\includegraphics[width = \columnwidth]{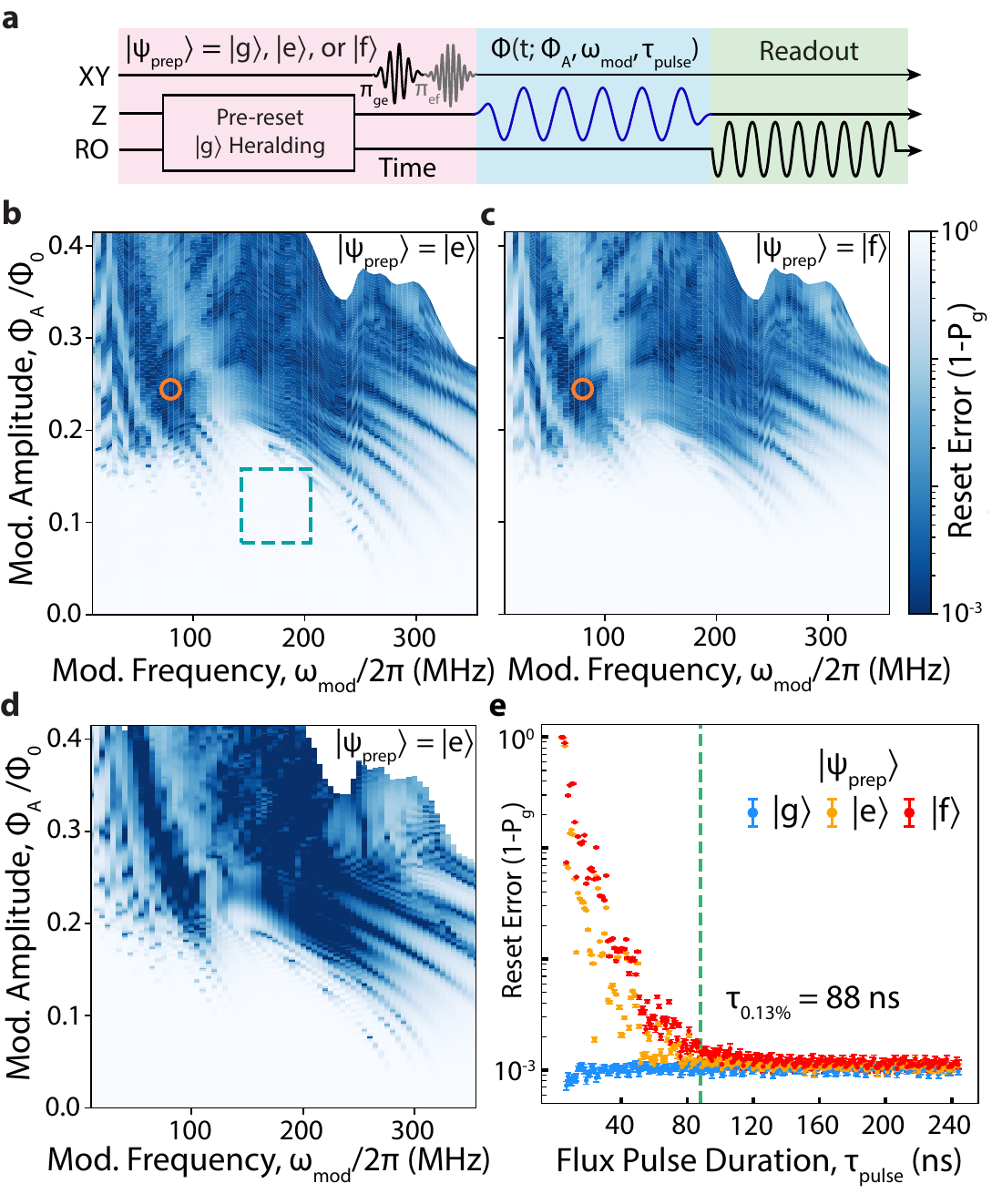}
\caption{\textbf{Calibration and characterization of unconditional reset.} \textbf{a}, Pulse sequence for characterizing reset operation. \textbf{b, c}, Reset errors obtained from scan of reset pulse parameters $\omega_\text{mod}$ and $\Phi_A$, for Q\textsubscript{1} prepared in $\ket{e}$ (\textbf{b}) and $\ket{f}$ (\textbf{c}). Dashed teal box indicates the scanned region for Fig. \ref{fig:LeakageReduction}b. \textbf{d}, Simulation of the measurement of \textbf{b} up to single-excitation subspace. \textbf{e}, Reset errors obtained by sweeping $\tau_\text{pulse}$ with 1 ns resolution, for $\Phi_\text{A} = 0.244\Phi_0$ and $\omega_\text{mod} = 80$ MHz (orange hollow circles in \textbf{b} and \textbf{c}). Error bars indicate 95\% confidence interval (t-test) with 30 samples, where each sample is obtained from 50,000 single-shot measurement.} 
\label{fig:UncondReset}
\end{figure}
The pulse sequence for characterizing a reset operation is provided in Fig. \ref{fig:UncondReset}a. We begin by pre-resetting Q\textsubscript{1} with a pre-calibrated reset pulse and heralding Q\textsubscript{1} in its ground state with high probability by using repeated rounds of projective measurement with a strict discrimination boundary~\cite{magnard2018fast}. Our ground state preparation has measured infidelity $1-P_{g, \text{HERALD}} \approx 0.02\%$, which is significantly smaller than the thermal population $\approx 0.1\%$ of the MMWG. We then apply a combination of $\pi_\textrm{ge}$ and $\pi_\textrm{ef}$ pulses to prepare Q\textsubscript{1} in its $\ket{g}$, $\ket{e}$, or $\ket{f}$ state, after which we apply the appropriate reset pulse, followed by readout of the qubit. The measured state populations are inverted using the calibrated confusion matrix. We define the reset error (infidelity) $\equiv 1 - P_g$ to be the population not discriminated to be in the ground state. 

Fig. \ref{fig:UncondReset}b and c shows the reset errors obtained from scanning the frequency and amplitude of a reset pulse of fixed length $\tau_\text{pulse} = 104$ ns, for prepared states $|e\rangle$ and $|f\rangle$. Thanks to the broadband nature of the MMWG, there are many frequency-amplitude pairs that could parameterize viable reset pulses with reset errors \textless $1\%$ for both prepared states. Simulation of the experiment of Fig. \ref{fig:UncondReset}b agrees very well with experimental data, replicating most of the significant qualitative features of the data, as shown in Fig. \ref{fig:UncondReset}d (see supplementary text \MakeUppercase{\romannumeral 5} for further details). Sets of curved features found in Fig. \ref{fig:UncondReset}b-d show where sidebands are dominantly interacting with the longer-lived bandedge modes of the MMWG passband. Additionally, the leftmost region of plot shows evidence of coherent dynamics due to the relatively slow tuning of the qubit at lower modulation frequencies through the more coherent bandedge modes. 

Fig. \ref{fig:UncondReset}e shows reset quality at a fixed frequency and amplitude as $\tau_\text{pulse}$ is varied. For a given frequency and amplitude, we define the time $\tau_\epsilon^{i}$ for reset of error $\epsilon$ such that for $\tau_\text{pulse} > \tau_\epsilon^{i}$, $1 - P_g < \epsilon$, for initial state $\ket{i} \in \{\ket{g}, \ket{e}, \ket{f}, ...\}$. We consider this definition to take into account time required for the depletion of photons in the MMWG after emission, which can re-excite the qubit upon spurious reflection from the MMWG taper. Fig. \ref{fig:UncondReset}e shows a frequency-amplitude pair that gives one of the fastest resets as characterized by $\tau_{\textrm{1\%}}^{e} = 44$ ns, $\tau_{\textrm{0.3\%}}^{e} = 51$ ns, and $\tau_{\textrm{0.13\%}}^{e} = 88$ ns when prepared in $\ket{e}$, with a steady-state population $P_\text{e}^\text{s.s.} = \textrm{0.11\%(2)}$ found from fit. This residual excitation population is close to the qubit excited state population when thermalized to the MMWG with its g-e transition frequency near the center of the passband. In addition, it is clearly demonstrated that the reset achieves simultaneous reset of multiple excited states, characterized by $\tau_{1\%}^{f} = 51$ ns, $\tau_{0.3\%}^{f} = 70$ ns, and $\tau_{0.16\%}^{f} = 88$ ns. 

Similarly, rapid reset of multiple excited states with a small residual excited state population of $\approx$ 0.2\% within $\approx$ 97 ns is also achieved when the Q\textsubscript{1} is biased at its lower sweet-spot (LSS) with a g-e transition frequency $\omega_{ge}^{LSS}/2\pi \approx 3.78$ GHz. This demonstrates the flexibility in choosing the idling frequency relative to the strongly damped auxiliary modes used for unconditional reset, while maintaining the speed and the ability to reset multiple excited states simultaneously. For the results of unconditional reset from the lower sweet-spot, refer to Appendix \ref{App:UncondReset}.

\begin{figure}[tbp]
\centering
\includegraphics[width = \columnwidth]{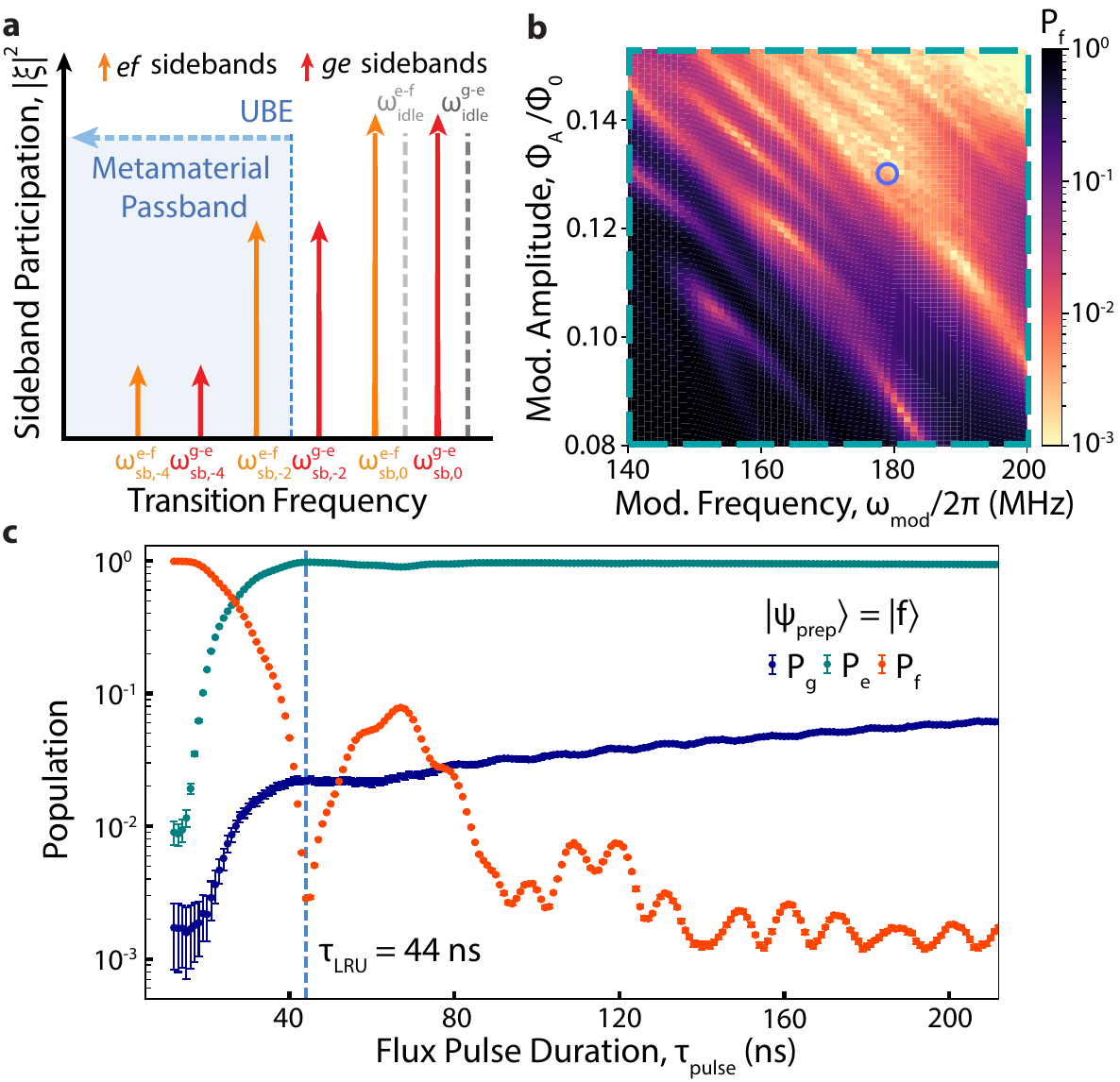}
\caption{\textbf{Calibration and benchmarking of LRU operation.} \textbf{a}, Sideband picture of the parametric modulation of Q\textsubscript{1} during LRU. \textbf{b}, $\ket{f}$ population after applying LRU implemented by parametric flux modulation, with two-dimensional scan of pulse parameters over the region indicated in Fig. \ref{fig:UncondReset} b and c (dashed teal box). \textbf{c}, Resulting $\ket{f}$ population obtained after applying LRU, where $\tau_\text{pulse}$ is swept with 1 ns resolution, for $\Phi_\text{A} = 0.13\Phi_0$ and $\omega_\text{mod}/2\pi = 179$ MHz (blue hollow circle in \textbf{b}). Error bars indicate 95\% confidence interval (t-test) with 30 samples, where each sample is obtained from 100,000 single-shot measurements. }
\label{fig:LeakageReduction}
\end{figure}

The sharp roll-off of the DOS at the upper bandedge of the MMWG relative to the transmon anharmonicity enables implmentation of leakage reduction units (LRUs), by selectively activating dissipation of $\ket{f}$ and higher excited states into the MMWG while maintaining the coherence of the g-e subspace.  Fig. \ref{fig:LeakageReduction}a shows the desired sideband configuration for implementing an LRU. An e-f sideband is placed within the passband of the MMWG, inducing radiative emission from $\ket{f}$ to $\ket{e}$, while the corresponding g-e sideband, greater in frequency by the transmon anharmonicity, remains outside the passband. 

It is desirable to minimize the time taken for the LRU so as to minimize the decoherence suffered by the g-e subspace due to relaxation and dephasing processes during leakage reset. This would generically require increasing the coupling of the e-f transition to the MMWG modes, which is accomplished in our parametric reset scheme by increasing the modulation amplitude for a given modulation frequency. However, depending on the amplitude of modulation, higher order sidebands of the g-e transition lying inside the MMWG passband may be of non-negligible participation, such as the fourth order sideband of the g-e transition as illustrated in Fig. \ref{fig:LeakageReduction}a, and will limit the fidelity of LRU. 

To determine viable parameter candidates for the LRU, we perform a more granular frequency - amplitude sweep with total flux pulse duration $\tau_\text{pulse} = 132$ ns. The explored parameter region is that indicated by the dashed teal box in Fig. \ref{fig:UncondReset}b, where we observe insignificant $\ket{e}$ decay and we expect the second order e-f sideband to just be entering the passband of the MMWG. In this measurement, we chose a smoother Gaussian filter parameter $\sigma = 3$ ns and buffer duration $\tau_\text{B} = 6$ ns in order to suppress spectral broadening of the second order sideband of the g-e transition that can induce unwanted additional radiative decay of the $|e\rangle$ state~\cite{lacroix2023fast}. The pulse sequence is identical to that of Fig. \ref{fig:UncondReset}a. As can be seen in Fig.~\ref{fig:LeakageReduction}b, multiple prominent regions of $\ket{f}$ relaxation are visible, which we believe correspond to interaction of the e-f transition with different modes near the upper bandedge of the MMWG.

The set of optimal LRU-parameters are found by trying to both maximize the $\ket{f}$ removal fraction and minimize the impact on remaining $\ket{e}$ population following LRU application (see Appendix~\ref{App:LRU} for further details in calibration of the LRU). For optimal modulation frequency $\omega_\text{mod}/2\pi = 179$~MHz and modulation amplitude $\Phi_\text{A} = 0.13\Phi_0$ (blue circle in Fig.~\ref{fig:LeakageReduction}b), we plot in Fig.~\ref{fig:LeakageReduction}c the measured final state g, e, and f probabilities as a function of LRU pulse length $\tau_\text{pulse}$ for an initial state of $\ket{f}$. 
We find the fastest LRU operation of 44 ns with a residual $\ket{f}$ population of 0.285(3)\%. Note that we define the time needed for LRU to be the time to the first local minimum of the $\ket{f}$ population, in contrast to the definition of the unconditional reset time which included the time required for the depletion of MMWG, for the following reasons. First, leakage errors are expected to occur infrequently ~\cite{miao2023overcoming}, hence the vast majority of the time we expect no emitted photons after the LRU. Any time spent waiting for waveguide ringdown in those cases serves no purpose and only incurs additional decoherence on spectator qubits. Second, in the event that leakage has occurred and the LRU has successfully returned the leaked population back to the computational subspace, a detected error in the next syndrome measurement is essentially inevitable. In this case, waiting for waveguide depletion to avoid additional computational errors is less valuable.

\begin{figure}[tbp]
\centering
\includegraphics[width = \columnwidth]{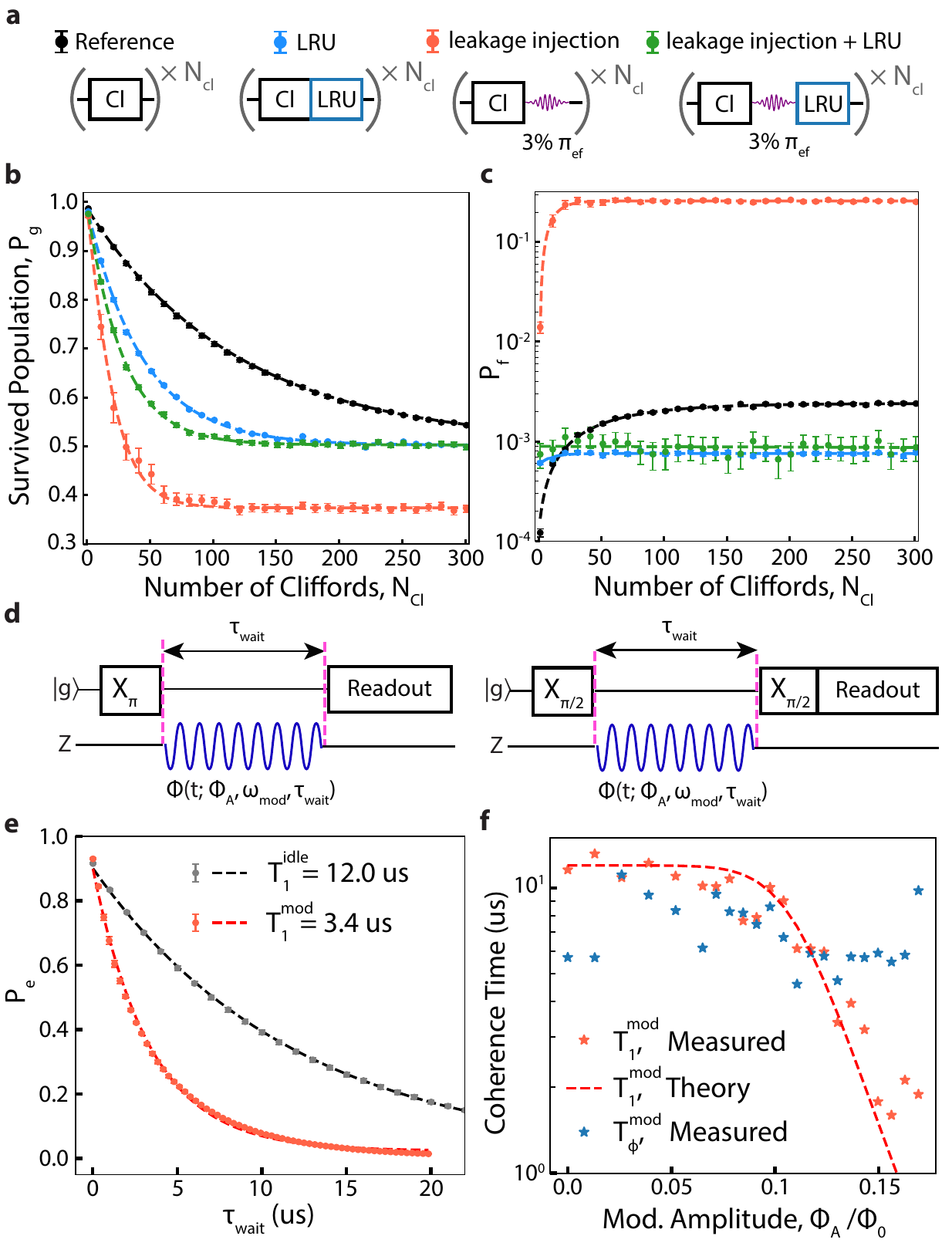}
\caption{\textbf{Randomized Benchmarking with LRU operation.} \textbf{a}, Primitive pulse sequences used for Randomized Benchmarking (RB) to assess the infidelity of the LRU as an idling gate (labelled ``LRU") and the efficacy of the LRU against realistic leakage errors (labelled ``leakage injection" and ``leakage injection + LRU"). Each sequence in the parentheses is repeated $\text{N}_\text{Cl}$ times to run an RB experiment. A $\pi_{ef}$ pulse with a reduced amplitude is used to artificially inject a leakage error of 3\%. Every LRU operation is followed by a pre-calibrated virtual-Z operation that cancels the phase accumulated during the LRU. \textbf{b, c}, Survived population (\textbf{b}, $P_g$) and accumulated leakage (\textbf{c}, $P_f$) after applying RB sequences with different numbers of Clifford gates ($\text{N}_\text{Cl}$). Dashed lines indicate fits to the model from~\cite{wood2018leakagerb}. Error bars indicate 95\% confidence interval (t-test) over 95 RB traces obtained over $\approx$ 25 hours. Each RB experiment is performed with 100 random Clifford sequences and averaged over 1,000 repetition. \textbf{d}, Pulse sequences used for measuring the relaxation coherence time, $T_1^\text{mod}$ (left), and the Ramsey coherence time, $T_2^\text{*,mod}$ (right), under continuous flux modulation during the waiting times $\tau_\text{wait}$. \textbf{e}, Comparison between relaxation coherence time without flux modulation ($T_1^\text{idle}$) and with flux modulation, using the parameters $\omega_\text{mod}/2\pi = 179$ MHz and $\Phi_\text{A} = 0.13\Phi_\text{0}$ as used in LRU. Dashed lines represent fits to an exponential decay profile. \textbf{f}, $T_1^\text{mod}$ and pure dephasing coherence time $T_{\phi}^{\text{mod}}$ measured at different $\Phi_\text{A}$, with $\omega_\text{mod}/2\pi = 179$ MHz. $T_{\phi}^{\text{mod}}$ is obtained from a relationship to Ramsey coherence time $(T_2^\text{*})^{-1} = (2T_1)^{-1} + T_\phi^{-1}$~\cite{ithier2005decoherence}. Red dashed curve represents the calculated $T_1^\text{mod}$ based on the sideband spectrum of the g-e transition during modulation and a simplified decay rate model within the MMWG passband provided in Appendix \ref{App:LRU}.}
\label{fig:LeakageReductionRB}
\end{figure}

With the calibrated LRU, we study its efficacy in removing leakage errors over long sequences and its decoherence impacts on the g-e subspace by utilizing randomized benchmarking (RB). For this analysis we consider four different RB primitive sequences: reference, interleaved LRU (``LRU"),  inject 3\% leakage after each Clifford gate (``leakage injection"), and leakage injection followed by LRU (``leakage injection + LRU"), shown schematically in Fig.~\ref{fig:LeakageReductionRB}a~\cite{lacroix2023fast, huber2024parametric}. 
Figures \ref{fig:LeakageReductionRB}b and \ref{fig:LeakageReductionRB}c compare the measured ground-state survival population and leakage accumulation, respectively, for each of these RB sequences. The RB curves are fitted to the leakage accumulation and the survival population models provided in Ref.~\cite{wood2018leakagerb}, which allows us to extract the steady-state leakage population of the RB sequences. When the leakage error rate per Clifford operation is significantly higher than the relaxation rates of the leakage states, the survived population $P_g$ is expected to drop below $50\%$ due to the buildup of a higher steady-state leakage population, as clearly observed in the case of ``leakage injection", where $P_f$ exceeds $20\%$. In contrast, we find that the use of the LRU effectively limits the accumulation of leakage significantly below that of the reference RB, resulting in a steady-state leakage population of $\approx 0.08\%$, which is close to the readout floor. Furthermore, the use of the LRU maintains a small steady-state leakage population of $\approx 0.09\%$, even when preceded by the injection of 3\% leakage. These results clearly demonstrate that the LRU is sufficiently effective in suppressing typical leakage error rates for single- and two- qubit gates in transmon qubits. 

Noting the sufficiently small leakage accumulation in both the ``Reference" and ``LRU" cases, we conduct a conventional interleaved RB (iRB) analysis~\cite{magesan2012efficient}, revealing a gate infidelity of 0.72(1)\%. To better understand the sources of this infidelity, we quantify the impact of flux modulation on the g-e subspace by measuring coherence with the pulse sequence shown in Fig.~\ref{fig:LeakageReductionRB}d. We observed a reduced relaxation coherence time $T_1^\text{mod} = 3.3$~$\mu$s during the flux modulation with the same parameters as used in the LRU operation, as shown in Fig.~\ref{fig:LeakageReductionRB}e. Additionally, we observe a clear reduction in $T_1^\text{mod}$ as the modulation amplitude is increased at a fixed modulation frequency, as illustrated in Fig.~\ref{fig:LeakageReductionRB}f. We attribute such reduction in relaxation coherence time to the activation of higher-order g-e sidebands within the MMWG passband, which serves as an unintended channel for photon emission into the MMWG. This interpretation is supported by the strong agreement with the numerically estimated $T_1^\text{mod}$ (red dashed curve), obtained from numerical calculation of sideband strengths and assuming constant emission rates within the MMWG passband. We also perform a Ramsey experiment with flux modulation activated to measure the decoherence time during the LRU operation and use this along with $T_1^\text{mod}$ to estimate the pure dephasing rate during modulation, $T_{\phi}^\text{mod}$. We find $T_{\phi}^\text{mod} = 3.7$ us, roughly uncorrelated with the modulation amplitude, as shown in Fig.~\ref{fig:LeakageReductionRB}f. 

From these independent measurements, we estimate that such decay through unwanted sidebands contributes infidelity of $\approx 0.23\%$ for the pulse parameters used for LRU. Including contribution from background (idling) decoherence rates, we estimate infidelity of 0.72(1)\% for the LRU restricted to the $g-e$ subspace, in good agreement with the measured infidelity. Further improvements in infidelity is expected for qubits with longer idling coherence time, since they allow for using a weaker modulation amplitude and longer pulse duration for LRU, suppressing decay through unwanted g-e sidebands. For further discussion on and analysis on LRU fidelity, refer to Appendix~\ref{App:LRU}.  

\label{Section:Discussion}
\textit{Summary and Outlook.}---We have demonstrated simultaneous unconditional reset of a transmon with residual excitation population below 0.13\% when prepared in $\ket{e}$ and below 0.16\% when prepared in $\ket{f}$, in 88ns. We additionally implement an LRU operation that resets a transmon's second excited state to 0.285(3)\% residual excitation in 44ns and has an overlap infidelity of 0.72(1)\% with the identity operation in the computational subspace.

Although the reset and LRU operations presented here are remarkably fast, further avenues for improvement are clear. The qubit-waveguide coupling could be increased further, as the qubit relaxation times are not yet Purcell-limited ($T_1^\text{Purcell-limit} > 1$ s). Narrowing the passband of the waveguide would increase the emission rate by increasing the photonic DOS in the passband. Such a narrowing of the passband would also help suppress spurious reflections at the input and output ports by improving the achievable impedance matching, and provide further Purcell protection during LRU by steepening the roll-off of the MMWG bandedges. Further reduction in reset time could be achieved by cutting down on the number of unit cells of the MMWG. This would allow the MMWG to ring down more quickly by reducing the round trip time, compensating for the increased group delay from a narrower passband. See Appendix~\ref{App:UncondReset} for more details. Regarding leakage reduction, use of MMWG with narrower passband that allows placing only one of the sidebands from the e-f transition would prevent emission via g-e sidebands through the MMWG passband ~\cite{ferreira2024deterministic}, providing better protection of g-e subspace and yielding lower infidelity for the LRU. Further discussion and analysis of LRU performance is presented in Appendix~\ref{App:LRU}.

The reported reset error level is understood to be limited here by the thermal population of the MMWG. We believe the temperature of the MMWG ($\approx 45$ mK) was elevated above the mixing plate temperature in our system due to an incidental overlap of one of our TWPA pumps with the MMWG passband combined with insufficient isolation allowing the pump tone to backpropagate from the TWPA to the device. Ameliorating this systematic issue as well as improving the MMWG tapering to its environment could help better reduce the residual excitation population after reset. See Appendix~\ref{App:ThermalPop} for further details.

While the unconditional nature of the reset obviates the need for the measurement of the transmon state, information, partial or otherwise, about the state of the transmon prior to reset can still be extracted by collecting the outgoing photonic field from any damped ends of the MMWG~\cite{besse2018single, reuer2022realization, ferreira2024deterministic}. This could, in the most ideal case, allow for conversion of leakage errors into erasure errors by detecting when leakage events occur.

Finally, while the demonstrated unconditional reset and LRU in principle simultaneously reset a few higher excited states including the third ($\ket{h}$) and the fourth excited state of the transmon, as their transition sidebands are expected to reside within the broad passband of the MMWG. Although it has not been characterized in this work owing to lack of sufficient coherence in these states due to their spectral proximity to the MMWG passband, (bare $\omega_{fh}$ is expected to be $\approx 7.12$ GHz, upper bandedge $\approx 7.1$ GHz) a device in which a MMWG possesses a lower frequency passband would allow for characterizing reset of such states.

\begin{acknowledgments}
We thank Mo Chen, Lucia De Rose, Piero Chiappina, Steven Wood, Srujan Meesala, David Lake, Jamison Stevens, and Matt Davidson for helpful discussions. We appreciate MIT Lincoln Laboratories for providing the traveling-wave parametric amplifier used in the microwave readout chain in our experimental setup. This work was supported through a grant from the Department of Energy (grant DE-SC0020152), and through a sponsored research agreement with Amazon Web Services. E.K. was supported by the New Faculty Startup Fund from Seoul National University, the POSCO Science Fellowship of POSCO TJ Park Foundation, the National Research Foundation of Korea (NRF) grant funded by the the Ministry of Science and ICT (MSIT) (No. RS-2024-00334169), and the LAMP Program of the National Research Foundation of Korea (NRF) grant funded by the Ministry of Education (No. RS-2023-00301976).
\end{acknowledgments}
\clearpage

\appendix

\renewcommand{\thefigure}{A\arabic{figure}}
\setcounter{figure}{0}

\section{Measurement Setup}
\label{App:Fab_Meas}
Our fridge wiring is identical to that of Fig. S6 of the Supplementary Material of~\cite{zhang2023superconducting}.

Our external control electronics wiring diagram can be seen in Fig. S1 here. It is similar to Fig. S5 of the Supplementary Material of~\cite{zhang2023superconducting}, except we have removed the 10dB of attenuation on the fast flux lines of subfigure a. to increase the dynamic range of our reset pulses, and the upconversion of our microwave XY signals is now handled by two Quantum Machines Octave units, obviating the need for some of the in-house filters, mixers, attenuators, amplifiers, and fast RF switches on those XY lines. Note that the readout upconversion and downconversion chain continue to use our previous in-house mixing configuration.

\section{Device Characterization}
\label{App:Device}
The circuit model of our device was identical to that of Supplementary Material section \MakeUppercase{\romannumeral 1} of~\cite{zhang2023superconducting} including fit values for the inductances and capacitances, except that we also included the readout resonators of each qubit as lumped-element LC resonators with center frequencies $\omega_{r\,i}$ fit from spectroscopy data of the device, inductances $L_r = 4.5$nH, and capacitances $C_{r\,i} = \frac{1}{\omega_{r\,i}^2 L_r}$. The readout resonators are capacitively coupled to each qubit with capacitance $C_{qr}$ and the unit cell of the MMWG with capacitance $C_{wr}$. We did not include parasitics for the readout resonators.

\section{Qubit Readout Calibration and Characterization}
\label{App:Readout}

\begin{figure}[tbp]
\centering
\includegraphics[width = \columnwidth]{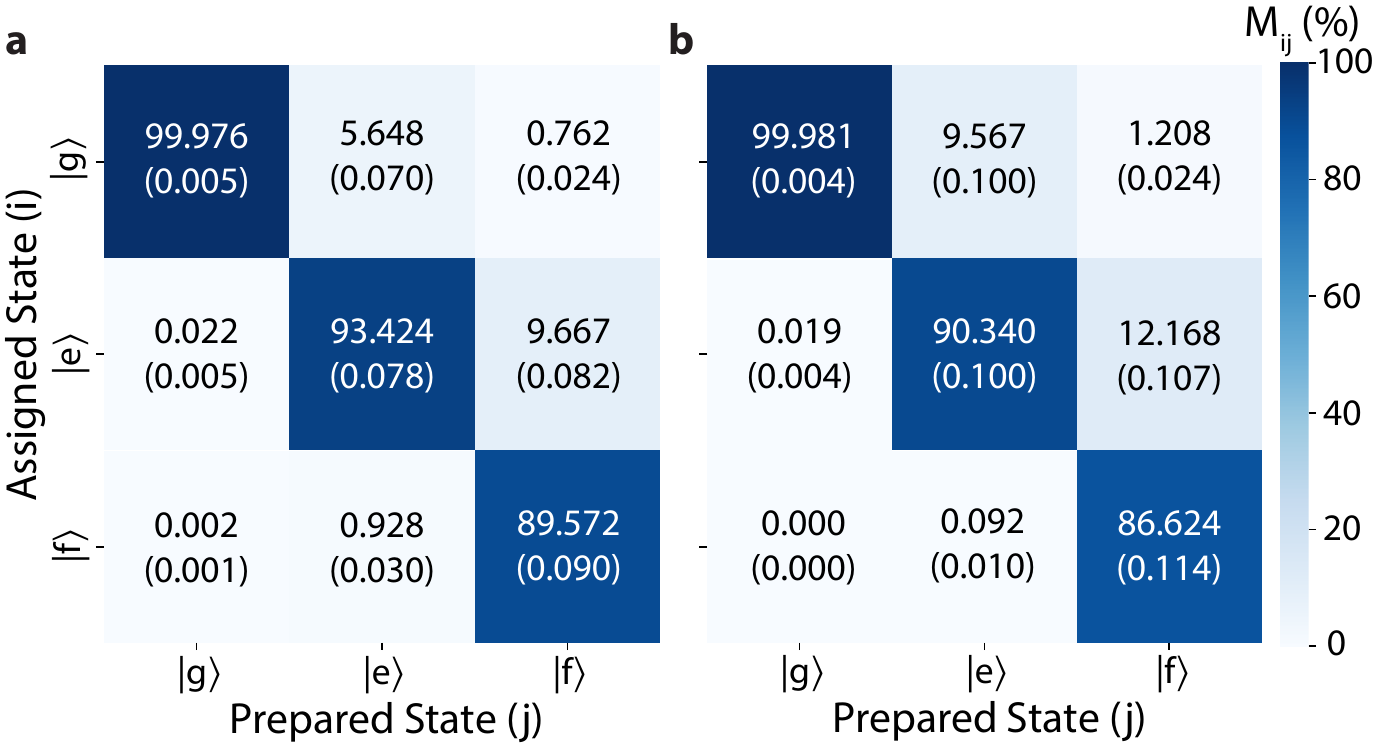}
\caption{\textbf{Estimated confusion matrix of single-shot readout.} Experimentally characterized readout confusion matrix $M_{ij}$ used for estimating $Q_1$'s population after applying the reset (\textbf{a}) and LRU operation (\textbf{b}), including the three lowest energy eigenstates of $Q_1$. The mean and standard error (in parentheses) of each matrix element are obtained over 100 repetitions, with each repetition consisting of 100,000 single-shot readouts.}
\label{fig:confusion}
\end{figure}

We perform three-level state discrimination of Q\textsubscript{1} with the standard transmon dispersive readout technique~\cite{sank2014fast, krantz2019aquantum}. Our input and output signal paths and amplification chain for readout is identical to that of Fig. S5-6 of the Supplementary Material of~\cite{zhang2023superconducting}. 

In this work, we send readout input microwave pulse at around the frequency of R\textsubscript{1} (the readout  resonator of Q\textsubscript{1}), $\omega_\text{drive} \approx \omega_{R1}$, with a 248ns-long square pulse envelope. We start the square envelope with a 20 ns kick with two times large amplitude for fast charging of readout resonator~\cite{mcclure2016rapid}, and convolved the envelope with a Gaussian kernel of 10 ns standard deviation. 

We measure temporal downconverted signals of both the microwave signals reflected (superscripted $R$) and transmitted (superscripted $T$) from the readout resonator and the MMWG Purcell filter using the OPX+, in order to improve SNR by $\approx$ 3dB. For each single-shot measurement (subscripted k), we collect the signals for 348ns, which is 100ns longer than the input readout pulse, considering the transient response of R\textsubscript{1}. The measured reflected and transmitted signals are then each mapped to a pair of real numbers ($I_k^R, Q_k^R$) and  ($I_k^T, Q_k^T$) by integrating them with pre-calibrated sets of integration weights, whose definition is given in the Supplementary Material of~\cite{krinner2022realizing}.

The four real numbers $u_{kj} \equiv (I_k^R, Q_k^R, I_k^T, Q_k^T)_j$ then characterizes the $k$-th single-shot readout of Q\textsubscript{1} prepared in state $\ket{j}$, $j \in \{g, e, f\}$. $\{u_{kj}\}$ yields three clusters of Gaussian distribution in a four-dimensional real space. Linear discrimination boundaries for real-time state discrimination are obtained by performing linear discriminant analysis (LDA)~\cite{fisher1936taxonomy, scikit-learn}. We first collect in total $3 \times 10^6$ single-shot readout data ($10^6$ for each prepared states). Instead of directly feeding the measurement data into LDA, we cluster the data with Gaussian Mixture (GM) clustering method, an unsupervised clustering technique assuming mixtures of Gaussian distributions. During tihs process, we use the \texttt{GaussianMixture} API from the Python package scikit-learn~\cite{magnard2018fast, scikit-learn}, and identify the corresponding state by selecting the state with the largest assignment probability. Next, we generate an artificially created ideal set of data $\{u_{kj, ideal}\}$, numerically drawn from the individual Gaussian distributions of each state. Finally, this ideal dataset is then fed into the LDA clustering algorithm by using \texttt{LinearDiscriminantAnalysis} API from scikit-learn, yielding linear discrimination boundaries. By using such artificially generated ideal samples instead of measured data, the process of determining the discrimination boundaries is significantly more resilient against various sources of errors including state preparation errors, $T_1$ relaxation, and measurement-induced state transitions.

We calibrate and characterize the readout by using readout confusion matrix $M \equiv \{M_{ij}\}$, where $M_{ij}$ is the probability of assigning state $\ket{i}$ when Q\textsubscript{1} is prepared in state $\ket{j}$ giving the following relationship,  
\begin{equation}
    \begin{gathered}
         \mathbf{P}^\text{A} = M \mathbf{P}
    \end{gathered}
\end{equation}

where $\mathbf{P} = (P_g, P_e, P_f)^T$ is the column vector of the transmon state population and $\mathbf{P}^\text{A} = (P_g^{A}, P_e^{A}, P_f^{A})^T$ is the column vector of the assigned state population. The measured populations present in this work, unless otherwise noted, are obtained by multiplying the inverse of the confusion matrix in order to account for the measurement errors. 
\begin{equation}
    \textbf{P} = M^{-1} \textbf{P}^\text{A}
\end{equation}

We aim at performing readout that is capable of accurately estimating residual excited state populations, while maintaining sufficiently large three-state assignment fidelity $F_{RO} \equiv (M_{11} + M_{22} + M_{33})/3$. We first run two-dimensional scan over readout pulse frequency and readout pulse amplitude and optimize for $F_{RO}$. Then, we increase readout pulse amplitude until the probability of assigning $\ket{e}$ or $\ket{f}$ when a sample is drawn from ideal distribution of $\ket{g}$ is sufficiently small $\approx 10^{-4}$. 

We start every measurement by heralding the $\ket{g}$ state using a stricter discrimination boundary with a target probability $P_\text{g, HERALD}^\text{target} = 0.9999$. This minimizes the effect of state preparation errors, and thus enables more accurate characterization of $M$. Note that negative populations that are often found in similar works may appear if the contribution from state preparation errors are not sufficiently excluded.

Fig.~\ref{fig:confusion}a shows the readout confusion matrices used for characterizing unconditional reset operations. We achieve readout floor $1-M_{gg} \approx 2\times 10^{-4} \approx 1-P_\text{g, HERALD}$ which is sufficiently small for estimating typical residual excited state population $\approx 0.1$\%, while maintaining high three-state readout fidelity of $F_{RO} \approx 94.3\%$. Fig.~\ref{fig:confusion}b shows the readout confusion matrices used for characterizing LRU operations. For this readout, we use a different pulse with higher amplitude in order to better estimate residual population of $\ket{f}$ states as characterized by $M_{fe} \approx 0.09\%$, while compromising $F_{RO}$ down to $\approx 92.3\%$ and maintaining similar readout floor of $\approx 2\times 10^{-4}$.

\section{Characterization of Metamaterial Waveguide as a Cold Bath}
\label{App:ThermalPop}

\begin{figure}[tbp]
\centering
\includegraphics[width = \columnwidth]{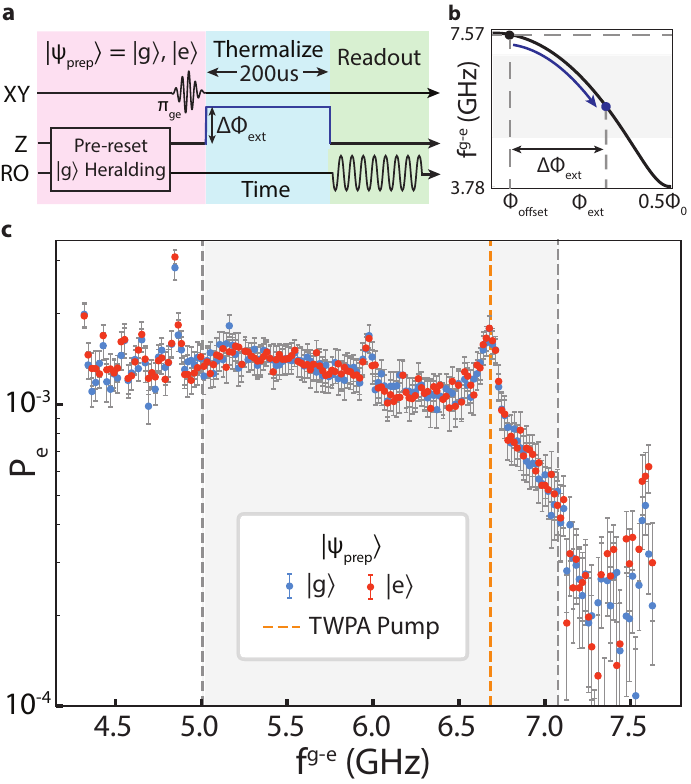}
\caption{\textbf{Characterization of the thermal excitation population at various g-e transition frequencies of Q\textsubscript{1}} \textbf{a, b}, Pulse sequence for characterizing thermal population. At each measurement, Q\textsubscript{1} is thermalized to its environment while a 200 us-long square flux pulse is applied, which brings g-e transition frequencies $f^{g-e}$ to various points. \textbf{c}, Residual Q\textsubscript{1} excitated state population $P_e$ after thermalization, for prepared states $\ket{g}$ (blue) and $\ket{e}$ (red). The range over which $f^{g-e}$ is swept includes the MMWG passband (gray shaded region) and USS frequency (rightmost point). The orange dashed line indicates the frequency of one the TWPA pumps (6.684 GHz) used for readout.}
\label{fig:thermal_pop}
\end{figure}

The thermal population of the MMWG is responsible for residual excited state population observed after applying the unconditional reset operations. We estimate the thermal population of the MMWG by thermalizing Q\text{1} to the MMWG for a sufficiently long time and reading out the excited state population of Q\textsubscript{1}. Fig.~\ref{App:ThermalPop}a and b shows the pulse sequence used for this process. We first prepare Q\textsubscript{1} in its $\ket{g}$ or $\ket{e}$ state using an appropriate combination of heralding into $\ket{g}$ and a $\pi_{ge}$. Then, the g-e transition  frequency of the Q\textsubscript{1} is tuned from 7.57 GHz to another frequency $f^{g-e}$ by sending a 200us-long square flux pulse. Finally, we record Q\textsubscript{1}'s residual excited state population via dispersive readout, as explained in the Appendix~\ref{App:Readout}.

The measured residual excited state population $P_e$ over different $f^{g-e}$ at which Q\textsubscript{1} idles is shown in Fig.~\ref{fig:thermal_pop}. The residual populations are almost the same for the two prepared states over most of the range, suggesting that the Q\textsubscript{1} reached a steady-state at each point and the residual excited state population can be understood as the thermal population of the MMWG. The thermal population of $\approx 0.1\%$ at 6GHz is corresponds to an effective temperature $T_\text{eff} \approx 43$ mK, which is consistent with the values reported in various works. There is an unexpected peak at 6.684 GHz due to insufficient isolation of one of the TWPA pump driven at 6.684 GHz. This suggests that the thermal populations could be further reduced if we isolate the signals and noise flowing from the TWPA to the output line further (note that we use 40 dB isolator), or avoid overlap between the MMWG passband and the frequencies of the TWPA pumps.

\begin{figure*}[tbp]
\centering
\includegraphics[width = \textwidth]{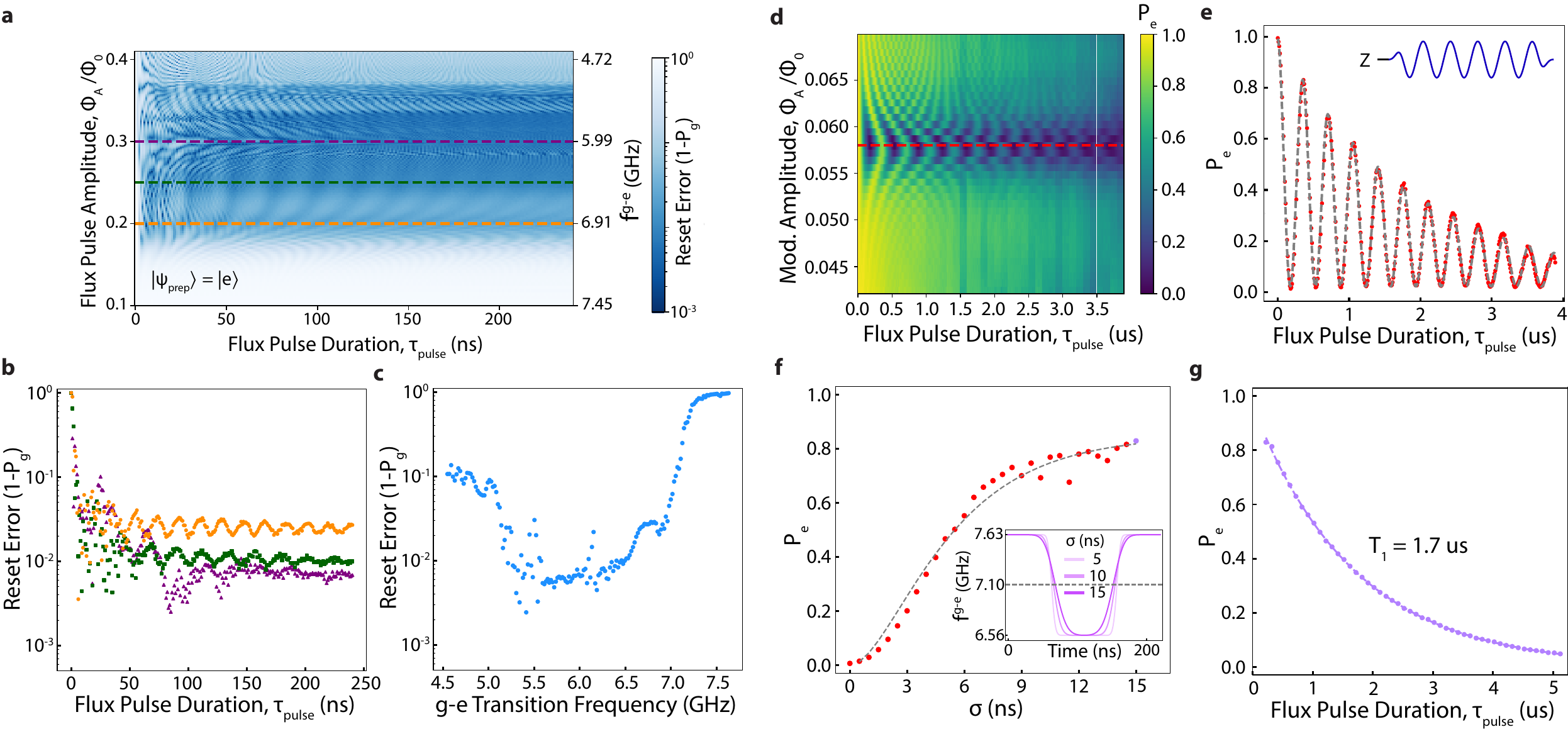}
\caption{\textbf{Reset with direct flux tuning, and probing presence and characterization of the long-lived resonances.} \textbf{a}, Quality of reset with direct flux tuning without AC modulation. Flux pulse amplitude $\Phi_\text{A}$ and flux pulse duration $\tau_\text{pulse}$ are swept. \textbf{b}, Cross-sectional view of \text{a} at various flux pulse amplitudes ($\Phi_\text{A} \approx 0.2\Phi_0$, $0.25\Phi_0$, $0.3\Phi_0$), showing the reset error over different pulse durations. \textbf{c}, Reset errors at $\tau_\text{pulse} = 240$ ns over different flux pulse amplitudes. (Cross-sectional view of \textbf{a} at $\tau_\text{pulse} = 240$ ns)   \textbf{d}, Probing the presence of one of the long-lived modes by coherent swap-interaction. Q\textsubscript{1} is prepared in $\ket{e}$, followed by a weak parametric flux modulation with $\omega_\text{mod} = 250$ MHz. Flux pulse amplitude and flux pulse duration are swept. \textbf{e}, Cross-sectional view of \textbf{d} (red dashed line). The measured populations (red dots) are fitted with a damped Rabi oscillation model (gray dashed line)~\cite{kosugi2005theoryofdamped}. \textbf{f}, Characterization of the long-lived mode by Landau-Zener tunneling experiments. The tuning rate of $f^{g-e}$ of the Q\textsubscript{1} when it crosses the frequency of the long-lived mode is varied by sweeping filter width $\sigma$ of the Gaussian kernel used for filtering the square flux pulses. (inset) The measured residual excited state populations (red dots) are fitted to a model (gray dashed line). \textbf{g}, Measured residual excited state population (purple dots) obatined by sweeping $\tau_\text{pulse}$, with $\sigma = 15$ ns. The fit (gray dashed line) characterizes the relaxation lifetime $T_1 = 1.7$ us of the long-lived mode. See appendix text for details of the Landau-Zener tunneling experiment.}
\label{fig:direct_tuning}
\end{figure*}

\section{Reset with Direct Flux Tuning and Long Lived Bandedge Resonances}
\label{App:DirectTuning}

Theoretically, direct tuning of transmon's transition frequencies within the MMWG provides even larger emission rate $\GammaoneD = \lambda_i^2g_\text{uc}^2 / J$ in the Markovian limit at the center of the passband~\cite{calajo2016atom, gonzalez2017markovian}. where $\lambda_i$ the relevant transition matrix element for the $i$-th state in the transmon annihilation operator. Consequently, it is expected that Q\textsubscript{1} would achieve a reset error of 0.1\% within 20 ns when prepared in $\ket{e}$, significantly faster that the reset demonstrated in the main text. However, as noted in the main text, we have chosen parametric flux modulation, in order to avoid adiabatic shelving of excitation into other modes that are coupled to the qubit which sets lower bound in achievable reset error and introduces non-Markovianity in the environment~\cite{spiecker2023two}.

Fig.~\ref{fig:direct_tuning}a-c shows the quality of reset by direct flux tuning. Despite the rapid initial decay of excited states, the reset error stops decreasing and reveals coherent dynamics dominating at longer pulse durations, over a wide range of g-e transition frequency during the flux pulse including the entire MMWG passband. 

Such phenomena can be understood by Landau-Zener-St\"uckelberg (LZS) interference with two passages; a fraction of excitation is transferred to long-lived resonances coupled Q\textsubscript{1} in the frequency collision during the rising edge of the flux pulse, and retrieved by the same mechanism during the falling edge of the flux pulse~\cite{shevchenko2010landau}. 

In order to test this hypothesis, we first probe the presence of a strongly-coupled long-lived resonance by activating a sufficient weak sideband of the g-e transition of Q\textsubscript{1} around the upper bandedge of the MMWG. We indeed observe damped Rabi oscillation when there is a weak sideband at $\approx 7.1$ GHz, which is a signature of two coupled coherent modes, evidenced by a chevron pattern and perfect fit to a model found in Fig.~\ref{fig:direct_tuning}d and e~\cite{kosugi2005theoryofdamped}.

We then proceed to quantitative comparison with predictions from the theory of LZS interference. We perform two-passage experiment, in which the time-dependent g-e transition frequency $\omega_{ge} (t)$ of Q\textsubscript{1} prepared in $\ket{e}$ crosses the frequency of the long-lived resonance (7.1 GHz) twice, by applying a square flux pulse (without AC modulation). Here, the tuning rate of $\omega_{ge} (t)$ when it crosses 7.1 GHz is controlled by the filter width $\sigma$ of the Gaussian kernel used for filtering the square flux pulse, as illustrated in the inset of Fig.~\ref{fig:direct_tuning} f. The remaining $\ket{e}$ population $P_{R}$ is measured over different $\sigma$ from 0 ns (no filtering) to 15 ns, for a fixed duration of $\tau_\text{pulse} = 220$, with buffers $\tau_\text{B} = 60$ ns at the start and the end. This choice of temporal parameters for flux pulses provides sufficient durations for the tested range of $\sigma$, during which Q\textsubscript{1} dumps most of its excitation into the MMWG when it is inside the MMWG passband, such that the $P_{R}$ is dominated by the population retrieved from the long-lived resonance during the second passage. Therefore, theoretical prediction for this measurement is given by the following set of equations, rather than the equations assuming maximum coherence for the resonances provided in~\cite{shevchenko2010landau},
\begin{equation}
\begin{gathered}
    P_{R} = P_\text{0} (1 - P_\text{dia})^2\\
    P_\text{dia} = e^{-2\pi \Gamma}\\
    \Gamma = \frac{g_\text{Q\textsubscript{1}-M}^2}{\text{d}\omega_{ge} / \text{d}t |_{\omega_{ge} = \omega_{M}}} 
\end{gathered}
\label{eq:landau-zener}
\end{equation}
where $P_\text{0}$ is the revival population in the limit of an infinitely fast frequency tuning, M denotes the aforementioned long-lived resonance with a 0-1 transition frequency $\omega_M = 7.1$ GHz, and $g_{Q\textsubscript{1}-M}$ is coupling between Q\textsubscript{1} and M. We find an excellent agreement between the measured data and fit to eq.(~\ref{eq:landau-zener}) with a strong coupling $g_\text{Q\textsubscript{1}-M} \approx 23.6$ MHz, as shown in Fig.~\ref{fig:direct_tuning}f. In addition, we measure relaxation lifetime $T_1 = 1.7$ us for the resonance M, by sweeping the flux pulse duration with $\sigma = 15$ ns, as provided in Fig.~\ref{fig:direct_tuning}g.

It is worth noting that the rate of adiabatic exchange of population $P_\text{adia} = 1 - P_\text{dia}$ provided in eq.(~\ref{eq:landau-zener}) is much greater for higher excited states of the transmons, due to their larger transition matrix elements. Consequently, this imposes challenges in resetting higher excited states with flux tuning with some adiabaticity that would be needed for unconditional reset of multiple excited states with narrow-banded auxiliary modes~\cite{mcewen2021removing}. In order to overcome this issue, we have chosen parametric flux modulation which provides continuous dissipation through sidebands while interleaving multiple avoided crossing passages with long-lived resonances coupled to transmons, which does not require knowing the presence of or careful characterization of the relevant long-lived resonances.

\section{Unconditional Reset}
\label{App:UncondReset}

\begin{figure}[tbp]
\centering
\includegraphics[width = \columnwidth]{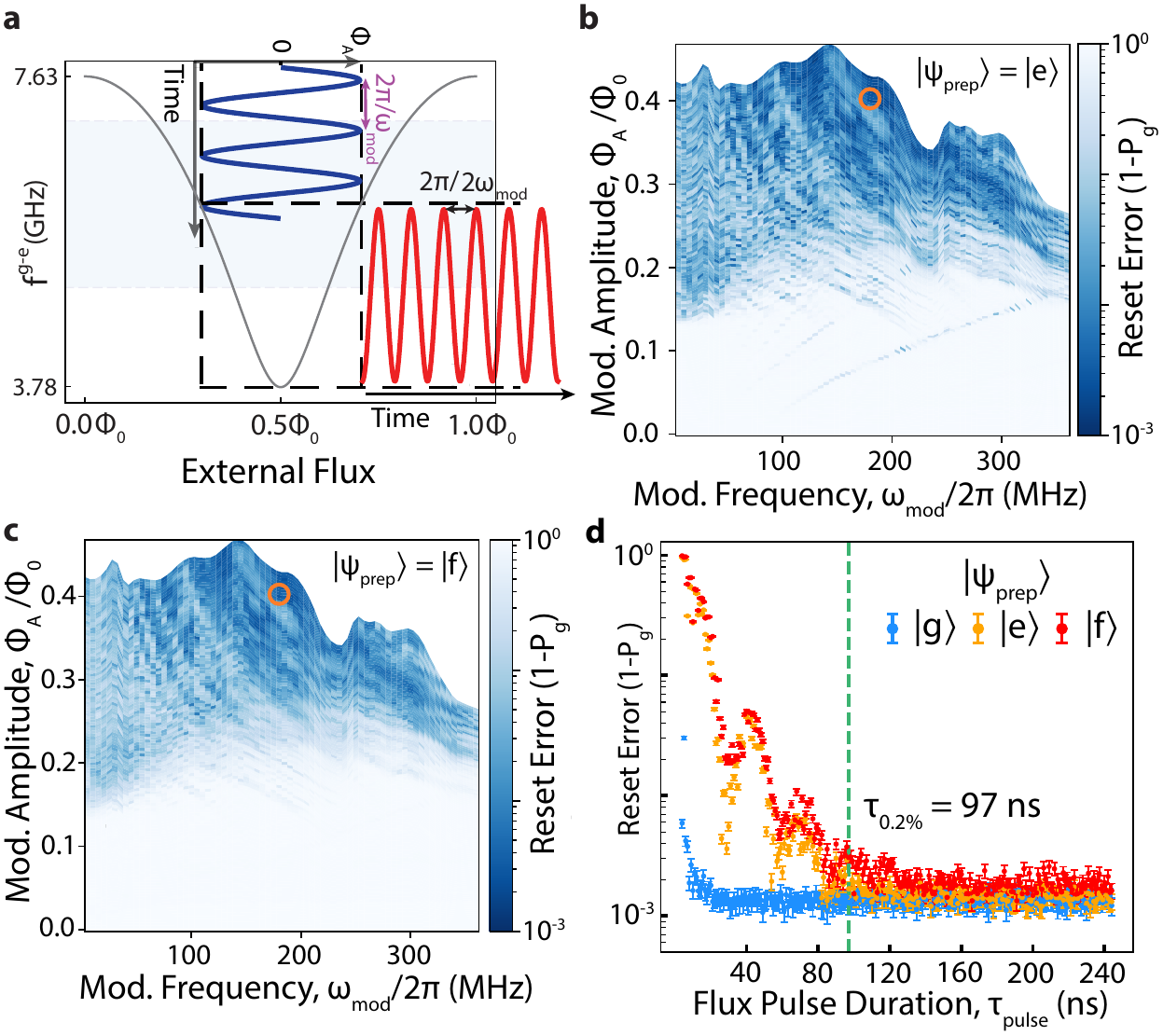}
\caption{\textbf{Uncondition reset from the lower sweet-spot (LSS)} \textbf{a}, Tuning curve of the flux-tunable transmon qubit centered at the LSS. $\Phi_0$ is the magnetic flux quantum. Parking the qubit at the LSS, where the qubit is first-order insensitive to the external flux, a flux modulation tone of frequency $\omega_{\textrm{mod}}$ (dark blue curve) modulates the qubit frequency at a frequency $2\omega_{\textrm{mod}}$ (red curve), similar to the case when the qubit is biased at its USS. The region shaded in pale blue denotes the extent of the MMWG passband. \textbf{b, c}, Reset errors obtained from scan of reset pulse parameters $\omega_\text{mod}$ and $\Phi_A$, for Q\textsubscript{1} prepared in $\ket{e}$ (\textbf{b}) and $\ket{f}$ (\textbf{c}). \textbf{d}, Reset errors obtained by sweeping $\tau_\text{pulse}$ with 1 ns resolution, for $\Phi_\text{A} = 0.4\Phi_0$ and $\omega_\text{mod} = 180$ MHz (orange hollow circles in \textbf{b} and \textbf{c}). Error bars indicate 95\% confidence interval obtained from bootstrapping.}
\label{fig:Lss_reset}
\end{figure}


\subsection{Unconditional reset from the lower sweet-spot}

In this appendix, we demonstrate the unconditional reset of a qubit when its idling frequency is below the passband of the MMWG. By positioning the MMWG passband above the qubit's idling frequency, we can achieve a lower residual population in the qubit after reset. This leverages the reduced thermal population at the higher frequency band of the MMWG, which is difficult to attain at the qubit's idling frequency~\cite{magnard2018fast, zhou2021rapid}.

We begin by biasing $Q_1$ at its lower sweet-spot (LSS), where we measure a g-e transition frequency of $\omega_{ge}^{LSS} = 3.78$ GHz and relaxation coherence time of $T_1 \approx 30$ us. As with the unconditional reset demonstrated in the main text for $Q_1$ biased at its upper sweet-spot, illustrated in Fig.~\ref{fig:Lss_reset}a, we apply a parameterized flux modulation pulse to activate sidebands within the passband of the MMWG. Fig.~\ref{fig:Lss_reset}b and c display the reset errors ($1-P_g$) as a function of the frequency ($\omega_\text{mod}$) and amplitude ($\Phi_\text{A}$) for a reset pulse with a fixed duration of $\tau_\text{pulse} = 104$ ns, for prepared states $\ket{e}$ and $\ket{f}$. 

Similar to the reset achieved from the USS, we find numerous frequency-amplitude pairs that enable unconditional reset with errors below 1\% for both prepared states. Fig.~\ref{fig:Lss_reset}d shows the reset errors obtained by varying the total pulse duration $\tau_\text{pulse}$ for the specific frequency-amplitude pair indicated by the orange circles in Fig.~\ref{fig:Lss_reset}b and c. Using the reset time definition from the main text, we find $\tau_\text{0.2\%}^e = 97$ ns and $\tau_\text{0.3\%}^f = 105$ ns, which are faster and as accurate as previously reported unconditional resets. This demonstrates the MMWG's capability for rapid and simultaneous reset of multiple excited states. Additionally, we observe a steady-state population of $0.15$\% from an exponential fit, corresponding to an effective temperature fo $\approx 28$ mK at the transition frequency of $3.78$ GHz. This indicates that the MMWG effectively provides a cold bath for reset.

Note that the population estimates shown in Fig.~\ref{fig:Lss_reset}d are acquired differently from other results in this work due to limited ground state preparation fidelity, which compromised the accuracy of the confusion matrix estimation. Instead, we determined the state populations by fitting IQ points from 100,000 repeated measurements to a mixture of Gaussian distributions using the \texttt{GaussianMixture} API from the Python package scikit-learn. The confidence interval for the estimated reset error was calculated through bootstrapping with 1,000 resamples.

\subsection{Simulation of reset dynamics}
We performed a Lindblad master equation simulation of the reset dynamics using the QuTiP~\cite{johansson2012qutip} python package to confirm the reset dynamics observed in measurement. Because of the overhead of quantifying the transfer function of our full flux lines for all the frequencies of interest in reset measurements, we did not attempt to use these simulations to obtain a quantitative fit of measured data, but only to verify that measured qualitative behaviour was expected in our best theoretical model of the device.

We peformed a full dissipationless circuit quantization procedure using the capacitance and inductance matrices of the device to define its Hamiltonian. Our circuit model was identical to that of Supplementary Material section \MakeUppercase{\romannumeral 1} of~\cite{zhang2023superconducting} including fit values for the inductances and capacitances, except that we also included the readout resonators of each qubit as lumped-element LC resonators with center frequencies $\omega_{r\,i}$ fit from spectroscopy data of the device, inductances $L_r = 4.5$nH, and capacitances $C_{r\,i} = \frac{1}{\omega_{r\,i}^2 L_r}$. The readout resonators are capacitively coupled to each qubit with capacitance $C_{qr}$ and the unit cell of the MMWG with capacitance $C_{wr}$. We did not fit the parasitics for the readout resonators.

We modeled dissipation at either end of the MMWG by defining two collapse operators on the modes corresponding to the final taper cells at either end of the waveguide. The associated decay rates for these collapse operators were defined to be the decay rate of the charge on the taper cell capacitor found by modeling the last taper cells as independent (that is, not coupled to the rest of the MMWG) LC resonators coupled capacitively to a 50 $\Omega$ resistor standing in for the environment and solving the equations of motion for the charge. More accurate modeling would incorporate more unit cells of the metamaterial waveguide.

Because we were not considering any drive on the system, and because of the excitation number preserving nature of the Jaynes-Cummings Hamiltonian, and the fact that the defined collapse operators only remove excitations from the system, we could restrict the total number of excitations in our system to be less than some finite value $N$. We chose $N$ = 1 because, given the total number of degrees of freedom in the MMWG-qubit-RO resonator system, $N$ = 2 was prohibitively expensive to simulate.

We simulated reset of Q1 for different flux pulses defined by different frequency-amplitude pairs with Q1 beginning at its upper sweet-spot and a single excitation in the eigenstate of the full Hamiltonian with the largest overlap with the bare Q1 eigenstate. Reset was simulated using a time-dependent term in the Hamiltonian that time-varied the frequency of Q1 according to the frequency and amplitude of the flux pulse, assuming the same Gaussian $\sigma = 1$ ns filtering of the flux pulse in simulation as was used in actual measurement and reset pulses of total length 104 ns. Results of the simulation can be seen in Fig.~\ref{fig:UncondReset}d.

\subsection{Evaluation of the demonstrated reset protocol}

\begin{figure}[tbp]
\centering
\includegraphics[width = \columnwidth]{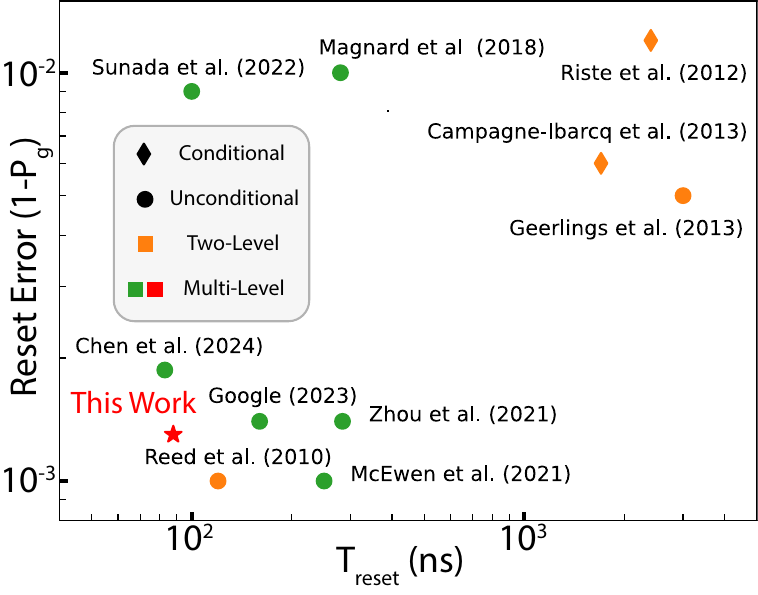}
\caption{\textbf{Comparison of experimentally demonstrated reset protocols.} Experimentally characterized reset errors (1-$P_g$) and the time to reach the quoted reset errors ($T_\text{reset}$) of various reset schemes reported in~\cite{reed2010fast, riste2012initialization, geerlings2013demonstrating, campagne2013persistent, magnard2018fast, zhou2021rapid, mcewen2021removing, sunada2022fast, miao2023overcoming}.}
\label{fig:Comparisons}
\end{figure}

In Fig.~\ref{fig:Comparisons}, we present a comparison of experimentally demonstrated reset protocols for superconducting qubits, evaluated by reset errors ($1-P_g$) and the time required to achieve these reset errors ($T_\text{reset}$)~\cite{reed2010fast, riste2012initialization, geerlings2013demonstrating, campagne2013persistent, magnard2018fast, zhou2021rapid, mcewen2021removing, sunada2022fast, miao2023overcoming, chen2024fast}. For conditional reset schemes, $T_\text{reset}$ includes the time needed for readout and the application of feedback on the qubits. For unconditional reset schemes, we quote $T_\text{reset}$ as either the pulse duration after which reset errors stay below the quoted reset errors or the total reset time including depletion of auxiliary modes.

\section{Leakage Reduction Units}
\label{App:LRU}

\begin{figure}[tbp]
\centering
\includegraphics[width = \columnwidth]{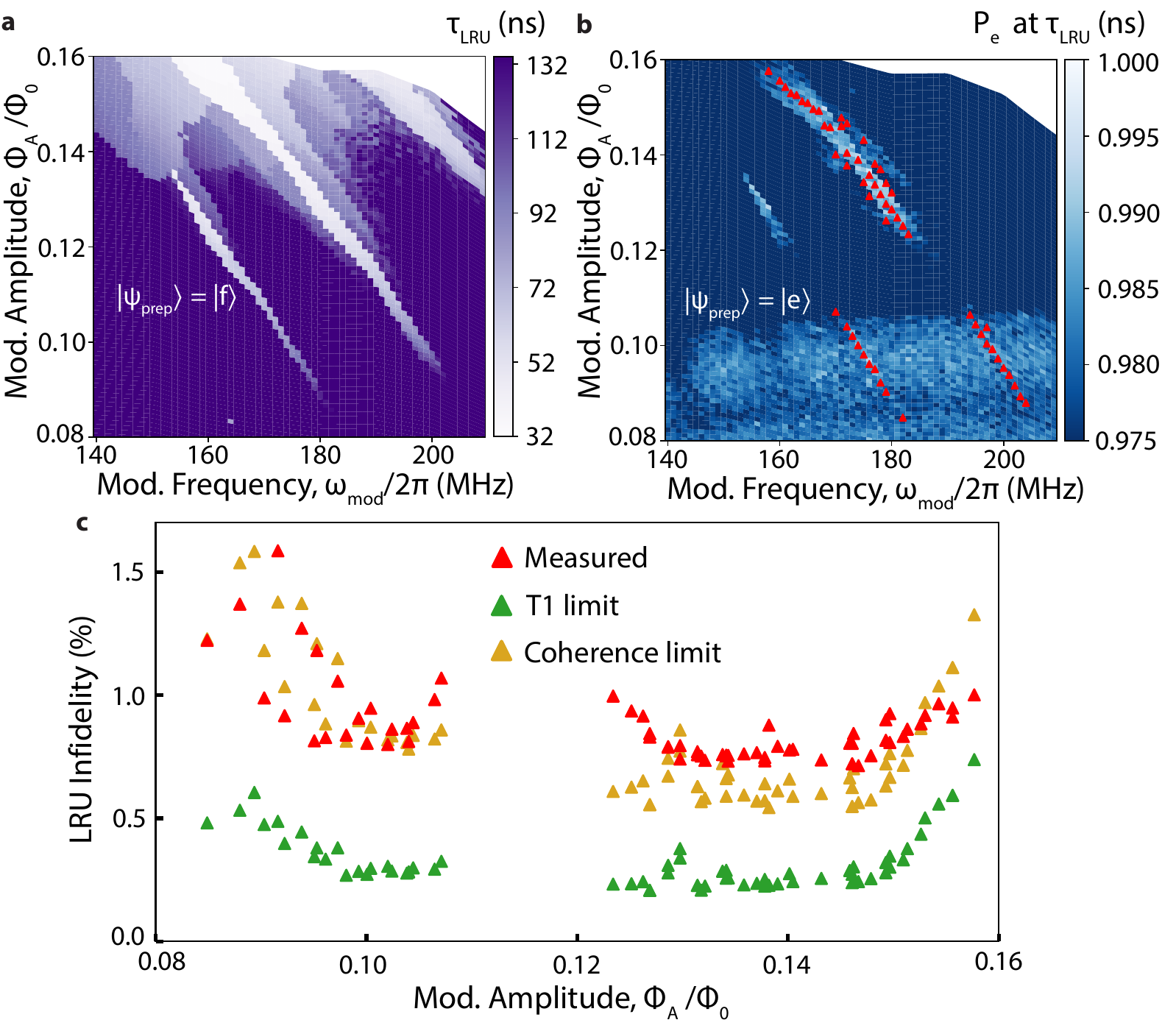}
\caption{\textbf{Calibration of LRU and impact of decoherence on infidelity in the g-e subspace.} \textbf{a}, $\tau_\text{LRU}$ over a range of ($\omega_\text{mod}$, $\Phi_\text{A}$) combination. $\tau_\text{LRU}$ is defined as the $\tau_\text{pulse}$ at which $P_f$ reaches the first local minimum when Q\textsubscript{1} is initially prepared in $\ket{f}$, obtained from a 3-dimensional sweep of ($\omega_\text{mod}$, $\Phi_\text{A}$, $\tau_\text{pulse}$). Note that the points without any local minimum are colored purple. \textbf{b}, Remaining $P_e$ population with a flux pulse with duration $\tau_\text{LRU}$, when Q\textsubscript{1} is initially prepared in $\ket{e}$. Red triangles indicate the parameter combinations for LRUs tested with iRB (\textbf{c}). \textbf{c}, Measured iRB infidelity of LRUs with different parameter combinations with $T_1$ and coherence limits. The coherence times ($T_1$ and $T_2^*$) are obtained from coherence measurements with the corresponding flux modulation tone applied continuously.}
\label{fig:lru_calibration}
\end{figure}

\subsection{Calibration of LRU parameters}

To characterize LRUs at each combination of parameters, we conduct a three-dimensional sweep of $\Phi_\text{A}$, $\omega_\text{mod}$, and $\tau_\text{pulse}$. For each frequency-amplitude pair, we determine the LRU duration $\tau_\text{LRU}(\omega_\text{mod}, \Phi_\text{A})$ by identifying the point where the population $P_\text{f}$ of $Q_1$, initially prepared in $\ket{f}$, reaches a local minimum, as shown in Fig.~\ref{fig:lru_calibration}a. This definition of LRU duration is the same as the definition provided in the main text. Fig.~\ref{fig:lru_calibration}b depicts the first excited state population $P_\text{e}$ after applying the LRU with the calibrated $\tau_\text{LRU}$, when $Q_1$ is initially prepared in $\ket{e}$. We find multiple LRU operation candidates, highlighted as the red triangles, where both a short $\tau_\text{LRU}$ and insignificant decay of $P_e$ are simultaneously achieved. Finally, using the interleave randomized benchmarking (iRB) technique~\cite{magesan2012efficient}, we estimate the gate fidelity of the LRU operation, followed by virtual-Z correction, when acting as an idling gate acting on the g-e subspace of $Q_1$. We have only included the points where leakage population ($P_f <$ 0.02) is sufficiently low toensure effective leakage removal, and where the population revival after reaching the first minimum is less than 20\%, excluding parameters where sufficient depletion of the MMWG modes is not achieved. 

At each candidate parameters, coherence limits of the gate infidelities (IFs) are calculated using the following formula derived from~\cite{pedersen2007fidelity},

\begin{equation}
\text{IF}(t_{\text{gate}}, T_{\phi}, T_1) \equiv 1-F_g \approx \frac{t_\text{gate}}{3} \left(\frac{1}{T_\phi} + \frac{1}{T_1}\right)
\label{eq:rb_coherence_limit}
\end{equation}

where $F_g$ represents the gate fidelity, and $t_\text{gate}$ denotes the duration of the gate. In estimating infidelities, we approximate  the LRU pulse as comprising two components: a buffer duration ($12$ ns) for the rising and falling edge of the envelope, during which the qubit is assumed to experience idling decoherence times ($T_1^{\text{idle}} \approx 12$ us and $T_\phi^{\text{idle}} \approx 7.3$ us), and the on-time ($\tau_\text{LRU} - 12$ ns), during which we assume the qubit experiences enhanced decoherence time measured under continuous flux modulation tone, as detailed in the main text. 

Fig.~\ref{fig:lru_calibration}c shows the distribution of LRU infidelities for the selected candidate parameters, along with calculated coherence limits and $T_1$ limits, which represent the contribution of $T_1$ to the overall coherence limit, as a function of modulation amplitude $\Phi_\text{A}$. The LRU infidelities align well with the coherence limit, suggesting that no other error channels play a significant role. At low $\Phi_\text{A}$, the LRU infidelities decrease as $\Phi_\text{A}$ increases, likely due to a reduced contribution from dephasing errors as $\tau_\text{LRU}$ shortens. However, at high $\Phi_{\text{A}}$, despite the shortening of $\tau_\text{LRU}$, the LRU fidelities begin to increase significantly due to increased $T_1$ decay rates during the sideband interaction. These two competing contributions result in a plateau of minimal infidelities in the range $\Phi_A \in [0.13\Phi_0, 0.15 \Phi_0]$, from which we select the LRU parameters used in the LRU operation discussed in the main text. The absence of candidates in the region $\Phi_A \in [0.11\Phi_0, 0.12 \Phi_0]$ is likely caused by a collision between the average qubit frequency of $Q_1$ ($0^{\text{th}}$ sideband) and a nearby resonance, such as two-level system defects (TLSs). This hypothesis is supported by the observed enhanced decay of $P_e$ across the entire modulation frequency $\omega_\text{mod}$, which minimally contributes to the average qubit frequency.

\subsection{Infidelity analysis of the LRU operation and ideal MMWG design for LRU}

\begin{figure}[tbp]
\centering
\includegraphics[width = \columnwidth]{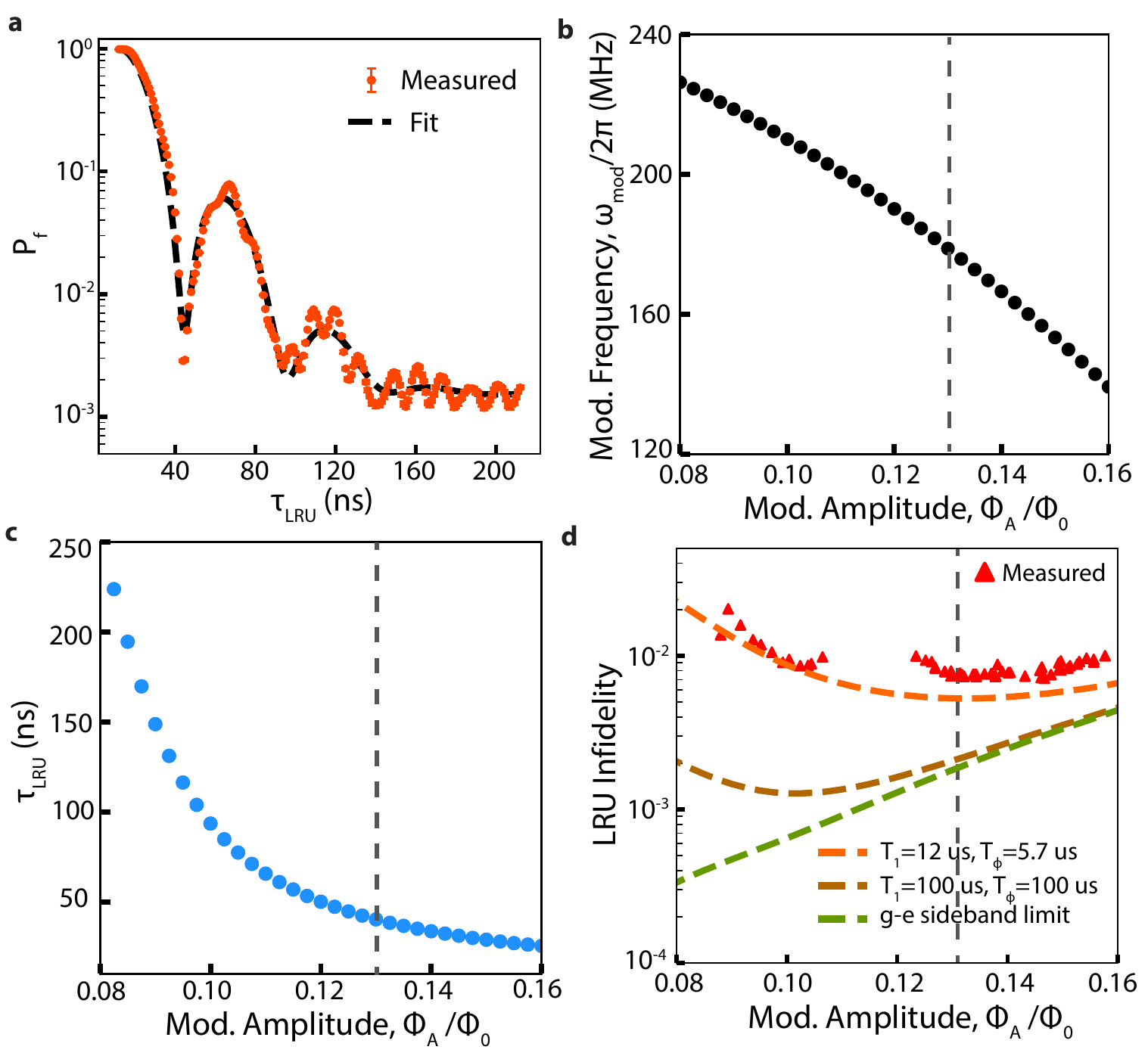}
\caption{\textbf{Approximation to single lossy-mode model and infidelity analysis for LRU operation} \textbf{a}, Resulting $\ket{f}$ population obtained after applying LRU, the same red curve in Fig.~\ref{fig:LeakageReduction}c. Black dashed line is the fit to the single lossy-mode model given by eq. (~\ref{eq:singlelossymode}), with $g_l/2\pi = 46.7$ MHz, $\omega_l/2\pi = 6.93$ GHz, $\kappa_l/2\pi = 17.6$ MHz, and $P_{s.s.} = 0.15\%$. \textbf{b}, Modulation frequencies $\omega_\text{mod}$ needed for different Modulation amplitudes $\Phi_\text{A}$ to locate the second lower ($m=-2$) sideband frequency of the $Q_1$ at the lossy mode's frequency $\omega_l$ to induce resonant interaction. \textbf{c}, $\tau_\text{LRU}$s estimated from the single lossy-mode model, at which $P_f$s (i.e., full-swap) reach their first local minima, for different modulation amplitudes $\Phi_\text{A}$s with modulation frequency $\omega_\text{mod}$s provided in \textbf{b}. \textbf{d}, Measured infidelities and infidelities estimated from eq. (~\ref{eq:infidelitymodel}), over different $\Phi_\text{A}$s. To understand the interplay between the optimal operating point of the LRU operation and the background decoherence rates of the qubits, we compare infidelity estimates with the measured decoherence rates (dashed red line), a case with improved coherence assuming $T_1 = 100$ us and $T_\phi = 100$ us (dashed brown line), and the case without any background decoherence labeled g-e sideband limit (dashed green line). Dashed gray lines in \textbf{b}, \textbf{c}, and \textbf{d} indicate the modulation amplitude $\Phi_\text{A} = 0.13 \Phi_0$ used for LRU operation shown in the main text, located at the plateau around the local minimum condition predicted by the model.} 
\label{fig:lru_model_infidelity}
\end{figure}

During the flux modulation with amplitude $\Phi_\text{A}$ and modulation frequency $\omega_\text{mod}$, the qubit-metamaterial interaction Hamiltonian can be written as the following, up to double-excitation subspace.

\begin{align*}
   &\hat{H}_\text{Q-MM}^\text{int} \approx \\ &\sum_k\sum_{j \in {1, 2}}\sum_m \lambda_j g_{k} \xi_m e^{i(\omega_k - \omega_m^{j-1, j})t} \ket{j-1}\bra{j}\hat{a}_k^\dagger + h.c.
\end{align*}

where $\ket{0}$, $\ket{1}$, and $\ket{2}$ indicates the transmon's three lowest energy eigenstates $\ket{g}$, $\ket{e}$, and $\ket{f}$, $\lambda_j \approx \sqrt{j}$ is the matrix element of the ladder operator for the transition between $\ket{j-1}$ and $\ket{j}$, $\omega_m^{j-1, j} = \overline{\omega^{j-1, j}(t)} + m \omega_\text{mod}$ is the frequency of the $m$-th order sideband of the $\ket{j-1}$-$\ket{j}$ transition, $\xi_m$ is the amplitude of the $m$-th sideband, $k$ denotes MMWG mode index, $\omega_k$ and $\hat{a}_k$ are the corresponding frequency and ladder operator, and $g_k$ is the coupling between the MMWG mode $k$ and the qubit. 

During an LRU operation, the coherent part of the dynamics of the e-f transition is dominated by the interaction with a single lossy waveguide mode located near the upper bandedge of the MMWG. In Fig.~\ref{fig:lru_model_infidelity}a, we show the dynamics of leakage population $P_f$ with varied LRU duration $\tau_\text{LRU}$ that is accurately fitted to eq. (~\ref{eq:singlelossymode}) describing a qubit interacting with a lossy waveguide mode denoted with index $l$,

\begin{equation}
\begin{gathered}
    H_\text{eff} = \ \begin{blockarray}{ccc}
    \ket{f0} & \ket{e1} \\
    \begin{block}{(cc)c}
        \Delta_{sb} - i\frac{\Gamma_{ef}}{2} & g_{sb} & \ \ket{f0}\\
        g_{sb} & - i\frac{\kappa_l}{2} & \ \ket{e1} \\
    \end{block}
    \end{blockarray}
    \\
    P_f^\text{sol}(t) = (1-P_{s.s.})\left| \frac{\alpha_-e^{\alpha_+ t} - \alpha_+e^{\alpha_- t}}{\alpha_- - \alpha_+}\right|^2 + P_{s.s.} \\
    \alpha_{\pm} = \frac{1}{2} \left( - b \pm \sqrt{b^2 - 4c}\right) \\
    b = \frac{\kappa_l + \Gamma_{ef}}{2} + i\Delta_{sb}, \ c = g_{sb}^2 + i\frac{\Delta_{sb}\Gamma_{ef}}{2} +\frac{\kappa_l\Gamma_{ef}}{4}
\end{gathered}
\label{eq:singlelossymode}
\end{equation}

where $H_\text{eff}$ is the effective Hamiltonian for the transmon-lossy MMWG mode's $\{\ket{f0}, \ket{e1}\}$ subspace via sideband interaction, $P_f^\text{sol} (t)$ is the analytical solution for the $Q_1$'s $\ket{f}$ state population during the sideband interaction with the initial condition $P_f^\text{sol}(0) = 1$, $\Delta_{sb} \equiv \omega_{-2}^{1,2} - \omega_{l}$ is the detuning between the the lossy mode $l$ and the closest sideband ($m = -2$) of the e-f transition, $g_{sb} \equiv \sqrt{2}g_{l}\xi_{-2}$ is the effective coupling through the sideband, $\kappa_{l}$ is the dampling rate of the lossy mode, $\Gamma_{ef} = 1/T_1^\text{ef}$ is the bare relaxation rate of the second excited state ($T_1^{ef} \approx 4.7 \mu $s), and $P_{s.s.}$ is the steady-state population. 

On the other hand, the information encoded in g-e subspace is affected by added decoherence by activating sidebands inside the passband of the MMWG. The total relaxation rates of transmon's states, $\Gamma_\text{tot}^{i-1, i}$, for the $\ket{i-1} - \ket{i}$ transition, can be approximated by summing the decay rates of each sidebands. Assuming a constant emission rate inside the MMWG passband, $\Gamma_\text{1D}^{i-1, i}(\omega) \approx \frac{\lambda_i^2g_{uc}^2}{J}$, and background relaxation rate, $\Gamma_{0}^{i-1, i} \equiv \frac{1}{T_1^{i-1, i}}$, outside the passband, we can use the following approximation.

\begin{equation}
    \Gamma_\text{tot}^{i-1, i} \approx \Gamma_0^{i-1, i} + \Gamma_\text{1D}^{i-1, i} \left(\sum_{\omega_m^{i-1, i} \in \text{Passband}} |\xi_m|^2\right)
    \label{eq:ge_relaxation_sideband}
\end{equation}

The dynamics and infidelity of LRU operations can be modeled using eq. (~\ref{eq:singlelossymode}) and eq. (~\ref{eq:ge_relaxation_sideband}). Fig. \ref{fig:lru_model_infidelity}a shows a fit to the measured $P_f$ for the model eq. (~\ref{eq:singlelossymode}), from which we obtain $g_l/2\pi = \frac{g_{sb}}{\sqrt{2}|\xi_{-2}|} = 46.7$ MHz, $\omega_l/2\pi = 6.93$ GHz, $\kappa_l/2\pi = 17.6$ MHz, and $P_{s.s.} = 0.15$\%. Based on this model, we find the LRU parameters $\omega_\text{mod}$  (Fig.~\ref{fig:lru_model_infidelity}b) and $\tau_\text{LRU}$  (Fig.~\ref{fig:lru_model_infidelity}c) at each modulation amplitude $\Phi_\text{A}$, such that $\Delta_{sb} = 0$ during the flux modulation and $P_f < 0.02$ after the LRU operation. 

With this model and using eq. (~\ref{eq:ge_relaxation_sideband}) and eq. (~\ref{eq:rb_coherence_limit}), we estimate infidelity of LRU operations for different flux modulation amplitudes by as shown in Fig.~\ref{fig:lru_model_infidelity}d (dashed orange line), from which a good agreement with the measurement LRU infidelity (red triangle) is found. In addition, we find that the modulation amplitude used in the main text (dashed gray line) is close to the optimal condition predicted by the model.

Further, we analyze the interplay between the background coherence of the qubit and the optimal condition for LRU operation by comparing the infidelity estimates with different coherence times assumed: $T_1 = 12 $ us, $T_\phi = 5.7$ us (this device, dashed orange line), $T_1 = 100$ us, $T_\phi$ = 100 us (dashed brown line), and no background decoherence (dashed green line, labeled g-e sideband limit). As longer background coherence tolerates longer LRU operation, further improvements in infidelity is achieved by using weaker modulation which suppresses higher-order sidebands inside the MMWG passband. Therefore, we expect LRU infidelity of $\approx 0.1$\% is achievable for qubits with improved background coherence ($\approx 100$ us).

\begin{equation}
    \label{eq:infidelitymodel}
\end{equation}


\section{Microwave Induced Leakage Reset}
\label{App:MicrowaveReset}
It is possible to perform leakage reset of a transmon dispersively coupled to a damped harmonic resonator by driving the transmon with a microwave charge drive at the difference frequency between the $\ket{f\,0}$ and $\ket{g\,1}$ states of the coupled system~\cite{pechal2014microwave, magnard2018fast}. This has the advantage of obviating the need for modulating the transmon frequency, thus avoiding the possibility of tuning the transmons through resonance with any other coupled modes in its spectrum (TLS, other qubits, etc.) and allowing leakage reset to even be performed on fixed frequency transmons.

Microwave leakage reset of a transmon capacitively coupled to a single harmonic mode is possible because the dispersive exchange coupling weakly dresses the bare $\ket{f\,0}$ and $\ket{g\,1}$ states in part with the $\ket{e\, 1}$ and $\ket{e\, 0}$ states respectively. The dressed states can then be coupled to one another via a charge drive on the transmon, as each dressed state contains some component of a state that differs from some component of the other only by a single excitation difference in the transmon part. If this driving is resonant with the energy splitting of the two states, the dressed $\ket{f\,0}$ and $\ket{g, 1}$ states will strongly hybridize and undergo Rabi oscillations, and if the harmonic mode is strongly damped, the hybridized state will then decay to the global $\ket{g\,0}$ ground state of the system. Because such a drive will in general not be resonant with any transitions out of the $\ket{g\,0}$ or $\ket{e\,0}$ states, the reset will only occur for leakage population in the second excited state, and the logical subspace of the transmon will be preserved.

By replacing the single harmonic mode with a metamaterial waveguide, and coupling the transmon to a single unit cell of the waveguide, the qubit couples to the many different harmonic eigenmodes of the metamaterial and can be reset through any of them. The Hamiltonian for such a system can be modeled as:
\begin{align}
        H =\; & \omega_q a^\dag a + \frac{\eta}{2} a^\dag a^\dag a a \\
         & + \; \sum\limits_{x=1}^{x=N} \omega_r b_x^\dag b_x + J (b_x b_{x+1}^\dag + b_x^\dag b_{x+1})\\
         & + \; g (a^\dag b_{x_0} + a b_{x_0}^\dag) \\ 
         & + \; \Omega \cos(\omega_d t) (a + a^\dag)
\end{align}
where $\omega_q$ is the transmon frequency, $\eta$ is the transmon anharmonicity, $x$ indexes a periodic linear chain of $N$, identical nearest-neighbor coupled resonators, comprising the metamaterial waveguide, each with frequency $\omega_r$ and nearest-neighbor coupling $J$, $g$ is the coupling between the transmon and the resonator indexed by $x_0$, and $\Omega$ is the strength of a drive at frequency $\omega_d$ on the transmon. The second line gives the bare metamaterial Hamiltonian, which can itself be diagonalized to give the bare eigenmodes of the waveguide in terms of the unit cell modes, and will also give the expression for $b_{x_0}$ in terms of the eigenmodes. When the microwave drive frequency $\omega_d$ is resonant with energy splitting between the $\ket{f\,0}$ state and the $\ket{g\,1}$ state, where the $0$ and $1$ indices refer to the $0$ and $1$ photon states of the single unit cell to which the transmon is coupled, it will induce an effective resonant coupling between the states $\tilde{g} = \frac{1}{\sqrt{2}}\frac{\eta \Omega g}{\Delta (\Delta + \eta)}$ where $\Delta \equiv \omega_q - \omega_r$~\cite{pechal2014microwave, magnard2018fast}.

To give a simple example of microwave reset using such a waveguide, it is possible to design a waveguide in a regime where $g / J$ is small while $g$ is on the order of at least a few times the FSR of the metamaterial $\sim (4J / N)$. In this regime, a microwave drive resonant with the $\ket{f\,0}$ state and the waveguide passband will turn on a resonant exchange interaction between the transmon and a bath of multiple waveguide modes, and the transmon's second excited state will roughly undergo exponential decay via photon emission into the waveguide at a rate $\Gamma \sim \frac{\tilde{g}^2}{J}$~\cite{calajo2016atom}. If the ends of the waveguide are well-damped to the environment, as with our tapered metamaterial waveguides, the photonic excitation will propagate away and the transmon-metamaterial system will decay to its global ground state. For $J / (2\pi) = 100$ MHz, $g / (2\pi) = 250$ MHz, $\eta / (2\pi) = -250$ MHz, $\Delta_{q-uc} / (2\pi) = -400$ MHz. With these parameters we get an FSR $\sim 8$ MHz, $\tilde{g} / (2\pi) \sim 25$ MHz, and $\Gamma_{1D} / (2\pi) \sim 6.5$ MHz. It is not necessary to operate in this particular regime, and the transmon leakage population should still be theoretically resettable even if the $\ket{f\,0}$ state is so weakly coupled that it sees well-defined modes of the waveguide, or so strongly coupled to individual modes that it undergoes non-exponential coherent dynamics with them as coupled system decays.

\vfill

\bibliography{bib}

\end{document}